\documentclass[ascmo, manuscript]{copernicus}
\usepackage{amsmath}
\usepackage{amssymb}
\usepackage{xcolor}
\usepackage{graphicx}
\usepackage{hyperref}

\nolinenumbers 

\definecolor{NR}{HTML}{00AEB3} 
\definecolor{YK}{HTML}{EC008C} 
\definecolor{CP}{HTML}{99479B} 
\definecolor{TC}{HTML}{00AEEF} 
\definecolor{LD}{HTML}{E98551} 

\newcommand{\NR}[1]{\color{black} #1 \color{black}}

\begin{document}
	
\title{Non-stationary GEV models for estimating design sea-states in a changing climate. Applications to offshore wind farms along the French coasts.}


\Author[1][nicolas.raillard@ifremer.fr]{Nicolas}{Raillard}
\Author[1,2]{Coline}{Poppeschi}
\Author[2]{Tessa}{Chevallier}
\Author[2]{Youen}{Kervella}
\Author[3,4]{Laurent}{Dubus}

\affil[1]{IFREMER, RDT, F-29280, Plouzané, France}
\affil[2]{France Energies Marines, Plouzané, France}   
\affil[3]{RTE, Puteaux, France}     
\affil[4]{World Energy \& Meteorlogy Council, Norwich, UK}            




\runningtitle{Non-stationary GEV models in a changing climate}

\runningauthor{N. Raillard, et al.}

\received{08/01/2026}
\pubdiscuss{} 
\revised{}
\accepted{}
\published{}


\firstpage{1}

\maketitle

\copyrightstatement{A \href{https://creativecommons.org/licenses/by-sa/4.0/deed.en}{CC-BY 4.0 International} public copyright license has been applied by the authors to the present document and will be applied to all subsequent versions up to the Author Accepted Manuscript arising from this submission, in accordance with the grant’s open access conditions.} 


\begin{abstract}
The rapid expansion of the French offshore wind sector requires a critical reassessment of structural durability in the face of evolving marine conditions driven by climate change. Traditional design methodologies, which rely on the assumption of stationary environmental conditions, are no longer adequate.

This study introduces a novel statistical framework to assess future changes in significant wave height by employing non-stationary Generalized Extreme Value (GEV) models applied to monthly maxima. This approach aims to reduce uncertainty and provide robust design tools adapted to the non-stationary conditions of the future. Based on CMIP6 climate models and reanalysis data, results reveal a projected trend towards a more pronounced seasonal contrast along the French Atlantic and English Channel coasts under future scenarios (SSP1-2.6 and SSP5-8.5), whereas the French Mediterranean Sea exhibits results that are more difficult to interpret, due to a weaker increase of extremes and large uncertainties (inter-model spread). Projections indicate more intense winters and calmer summers, along with a shift in the seasonal cycle. Overall, the multi-model ensemble suggests an increase in the design levels for extreme sea states.

The research concludes by defining a new methodology for calculating an equivalent design level over the structure's operational lifespan. This tool is deemed essential for ensuring the resilience and economic viability of future offshore wind farms in a changing climate.
\end{abstract}





\introduction 

The French offshore wind sector is undergoing rapid expansion, with a projected cumulative capacity of 3.6 GW expected by the end of 2027, driven by the commissioning of seven major offshore projects. This growth occurs in a context that is both innovative and promising, yet increasingly competitive.

However, the long-term durability and reliability of these offshore structures critically depend on their ability to withstand evolving marine conditions over operational lifespans that typically exceed two decades. Climate change is altering the dynamics of the marine environment, with notable trends in wave patterns, wind speeds, and the frequency and intensity of extreme weather events (\citet{amlashi2024}). These changes introduce substantial uncertainties into the design conditions of offshore wind farms (\citet{barkanov_evolution_2024}), particularly in regions exposed to high wave energy and strong winds; areas that are also the most favorable for energy production (\citet{susini_climate_2022}).

Metocean studies are essential for understanding and forecasting the environmental conditions that offshore structures will face throughout their service life. In particular, extremes in significant wave height, total sea level, and wind speed are critical design parameters that directly influence structural performance, safety, and operational efficiency (\citet{slater2021}). Significant wave height, representing the average height of the highest third of waves, is a key factor in determining wave loading on structures and plays a central role in fatigue analysis, foundation stability, and the overall design of wind turbine components (\citet{zhang_economic_2019}).

Climate-induced changes in extreme wave conditions may lead to increased wave loads. Underestimating these loads can result in structural damage, premature fatigue, or even catastrophic failure (\citet{schloer2013}). Traditionally, offshore wind turbine design has relied on statistical analyses of historical metocean data, assuming stationary environmental conditions. However, the non-stationary nature of climate change challenges this approach. Sea level rise, shifting storm tracks, and intensifying wave conditions can significantly affect wave loading, fatigue life, and structural integrity (\citet{chella2012}).

Neglecting these evolving trends may lead to under-designed structures, reduced operational lifespans, and increased maintenance costs,ultimately compromising the economic viability and safety of offshore wind projects. It is therefore imperative to assess the impact of climate change on sea state conditions to inform the design criteria of offshore wind farms and other coastal infrastructure.
This research aims to address this challenge by suggesting a new methodology to better investigate projected changes in significant wave height. The final objective is to provide engineers, designers, and stakeholders with updated tools that account for non-stationary environmental conditions, ensuring that offshore wind farms remain resilient and cost-effective in a changing climate.

This paper explores extreme wave conditions along French coasts using recent climate models coupled with advanced statistical approaches. We propose a novel methodology based on monthly maxima and long-term non-stationary models explained in the third section, modeling of extreme values. Then in the fourth section, results, this framework is then used to highlight changes in the return period for extreme sea states up to 2100. And finally, the predictions of extreme waves across the three French coasts are presented, and the models are thoroughly discussed and documented in the fifth section, discussion.

\section{Sea-state data}

\subsection{Sea-state data}
\subsubsection{CMIP6 Climate models}
Climate models are an essential tool for assessing future environmental responses, with the Coupled Model Intercomparison Project Phase 6 (CMIP6) representing the latest generation available from the IPCC (AR6, \citep{ipcc_ar6_europe}). The wave data used in this study are specifically derived from a numerical wave model (WaveWatch III) that was forced by eight General Circulation Models (GCMs) from the CMIP6 ensemble (listed in Table~\ref{tab:model_description}). More details about this data can be found in the article of \citet{meucci2024}. To align with the AR6 framework, we use the period 1985-–2014 as the baseline, and 2081--2100 for future projections. This dataset includes two climate change scenarios: SSP1-2.6 and SSP5-8.5. The wave data have a 3-hour frequency output and a 0.5$\degree$ spatial resolution.

\subsubsection{Reanalyses} 
Reanalyses are necessary to evaluate the quality of climate models over the past period, called "historical period", and to make bias-adjustment for future projections when necessary.
In the English Channel and the Atlantic Ocean, the HYWAT\footnote{https://doi.org/10.17183/REJEUX\_HYWAT} reanalysis was used to make this bias correction. The HYWAT reanalysis extends from latitudes between 43\degree N and 52\degree N and longitudes between 7\degree W and 5\degree E in a non-structured grid. Data is hourly with a spatial resolution from approximately 500 meters on the coast to a few kilometers offshore. The reanalysis is based on the configurations of HYCOM and Wavewatch III with a parametrization corresponding to TEST 471 (\citet{michaud2024}). HYWAT reanalysis is forced by the ERA5 atmospheric reanalysis for the wind and atmospheric pressure fields (\citet{hersbach2020}). HYCOM currents, water levels and surges (\citet{jourdan2020}) are also provided to the Wavewatch III model every 10 minutes.
In the Mediterranean Sea, the MED-WAV reanalysis\footnote{https://doi.org/10.25423/cmcc/medsea\_multiyear\_wav\_006\_012} was used to make the bias correction. The MED-WAV reanalysis has two grids. The coarse grid covers the North Atlantic Ocean from 75\degree W to 10\degree E (1/6\degree resolution) and from 70\degree N to 10\degree S while the fine grid covers the Mediterranean Sea from 18\degree W to 36\degree E and from 30\degree N to 46\degree N (1/24\degree resolution). Data is hourly with 1/24\degree resolution. The reanalysis is based on the WAM 4.6.2 model. MED-WAV reanalysis is forced with daily averaged currents from Med-Physics and with 1-h, 0.25$\degree$ horizontal-resolution ERA5 reanalysis 10 m-above-sea-surface winds from ECMWF.

\begin{figure}[ht]
\centering
\includegraphics[width=0.5\textwidth]{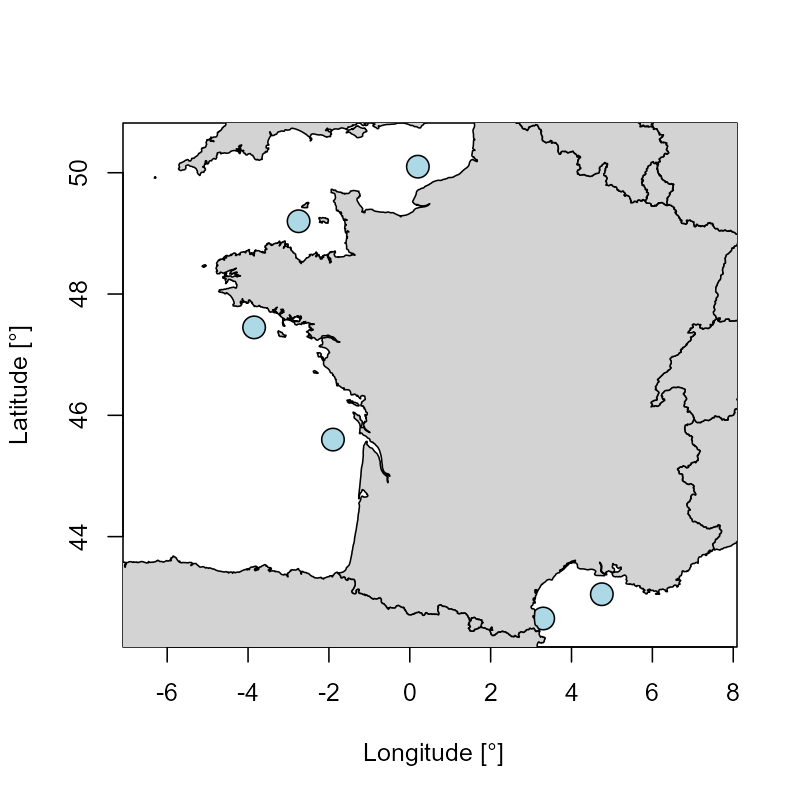}
\caption{Locations of the Selected Representative Point (SRP) along the French coasts.}
\label{fig:SRP-MAP}
\end{figure}

\begin{table}[ht]
    \centering
    \begin{tabular}{|l|l|l|}
        \hline
        Models	& Provider &	Reference\\
        \hline
ACCESS-CM2	&ACCESS (Australia)& \citet{Bi2020}\\
AWI-CM-1-1-MR	&AWI (Germany)&	\citet{Semmler2020}\\
CMCC-CM2-SR5	&CMCC (Italy)&	\citet{Lovato2020} \\
EC-Earth3	&EC-Earth-Consortium (Europe) &	\citet{EECE2019}\\
IPSL-CM6A-LR	&IPSL (France)&	\citet{Boucher2018}\\
KIOST-ESM	&KIOST (Korea)&	\citet{Pak2021}\\
MPI-ESM1-2-LR	&MPI (Germany)&	\citet{Wieners2019}\\
MRI-ESM2-0	&MRI (Japan)& \citet{Yukimoto2019}\\
\hline
    \end{tabular}
    \caption{List of the wind/waves climate models used in this study.}
    \label{tab:model_description}
\end{table}

\begin{table}[ht]
\centering
    \begin{tabular}{|l|l|l|}
        \hline
        Name	& Latitude (WGS84) & Longitude (WGS84)\\
        \hline
Eastern English Channel & 50.10 & 0.20 \\
Western English Channel & 49.20 & -2.75 \\
Northern Atlantic seafront & 47.45 & -3.85 \\
Southern Atlantic seafront & 45.60 & -1.90 \\
Western French Mediterranean & 42.65 & 3.30 \\
Eastern French Mediterranean & 43.05 & 4.75 \\
\hline
    \end{tabular}
    \caption{Seafront representative points coordinates.}
    \label{tab:seafront_coords}
\end{table}

\subsection{Bias correction}

We applied the Cumulative Distribution Function transform \textbf{CDF-t} method (\citet{michelangeli_probabilistic_2009}) to both correct the bias and to downscale large-scale GCMs data to the specific locations, using local-scale reanalysis as the reference. This correction was applied at six offshore locations in this study (situation in Figure~\ref{fig:SRP-MAP} and coordinates in Table~\ref{tab:seafront_coords}), selected as representative sites for future offshore wind farms across the French seafronts.

\begin{figure}[ht]
\centering
\includegraphics[width=0.75\textwidth]{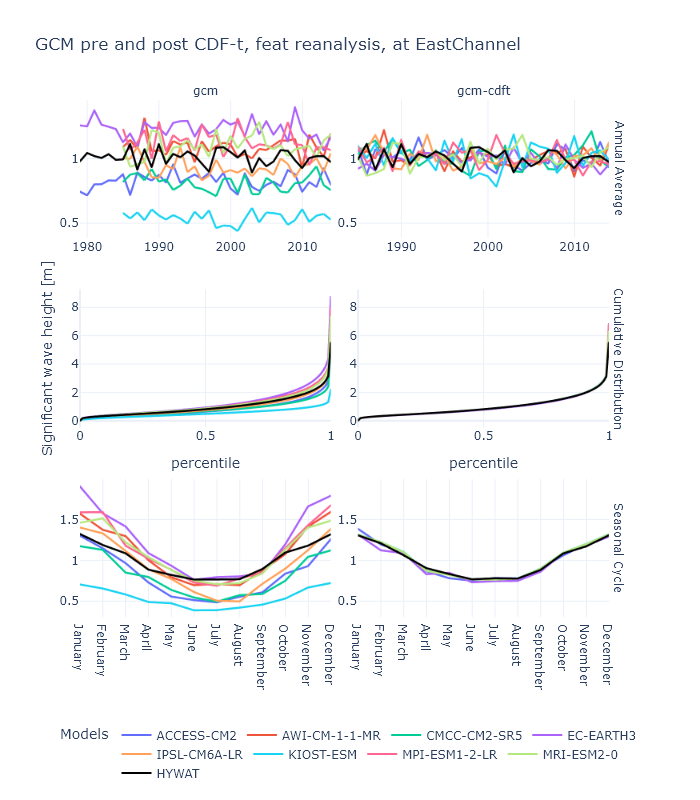}
\caption{CDF-t validation over the Eastern Channel point.}
\label{fig:cdft_validation}
\end{figure}

Figure \ref{fig:cdft_validation} illustrates the impact of this correction through the annual average, the cumulative distribution function (CDF), and the monthly average. The correction significantly reduces the dispersion of the models' annual average relative to the reanalyses. Although the CDF-t preserves the trend across all quantiles, adjusting only the rank and distribution properties (not the mean), the dispersion in the seasonal cycle is also notably diminished. This latter improvement is achieved because the bias correction was applied separately to each month.

\section{Modeling of extreme values}

\subsection{Models for annual maxima}

Following \citet{Coles2001}, we model the \textbf{yearly maxima} using a \textbf{GEV} distribution: $$M_n \sim \text{GEV}(\mu, \sigma, \xi),$$ where $\mu$ is the location parameter, $\sigma$ a scale parameter and $\xi$ a shape parameter.

From a practical point of view, the data is split by year, model and scenario, and the \textbf{GEV} distribution is fitted on the yearly maxima of each group, using a maximum-likelihood approach.

Once the model is fitted, we are able to go beyond the range of observed values, and estimate any quantile of the distribution of the yearly maxima, what is referred to as \emph{return level} in the literature. In figure~\ref{fig:compar_RL}, we show the estimations of such quantities from the fitted model, for each location and under each scenario. We have also checked individually that the fitted model fit well the data by looking at QQ-plots (not shown here).

\subsection{Models for monthly maxima}

\subsubsection{Methodology} 

We will follow the methodology of \citep{Reinert2021, Roustan2022, Cheynel2025}, briefly described hereafter. When taking the maxima on yearly blocs, a lot of data is not used, which results in large uncertainties in the fitted parameters due to the paucity of data. However, when using smaller blocs size, e.g. monthly blocs, we have to take into account the seasonality of the data, because the monthly maxima cannot be assumed to be identically distributed (see e.g. \cite{trasch2023}). 

\begin{equation}
\label{non-stat}
    M_{\text{m,n}} \sim \text{GEV}(\mu = f_1(\text{m}),\sigma = f_2(\text{m}),\xi),
\end{equation}
where \textit{m} is the month of the year. Because estimating $\xi$ is difficult (see \cite{Coles2001}), it is assumed to be constant for all the months.

\subsubsection{Computation of return levels} \label{retlev-monthly}

\begin{figure}[ht]
\includegraphics[width=\textwidth]{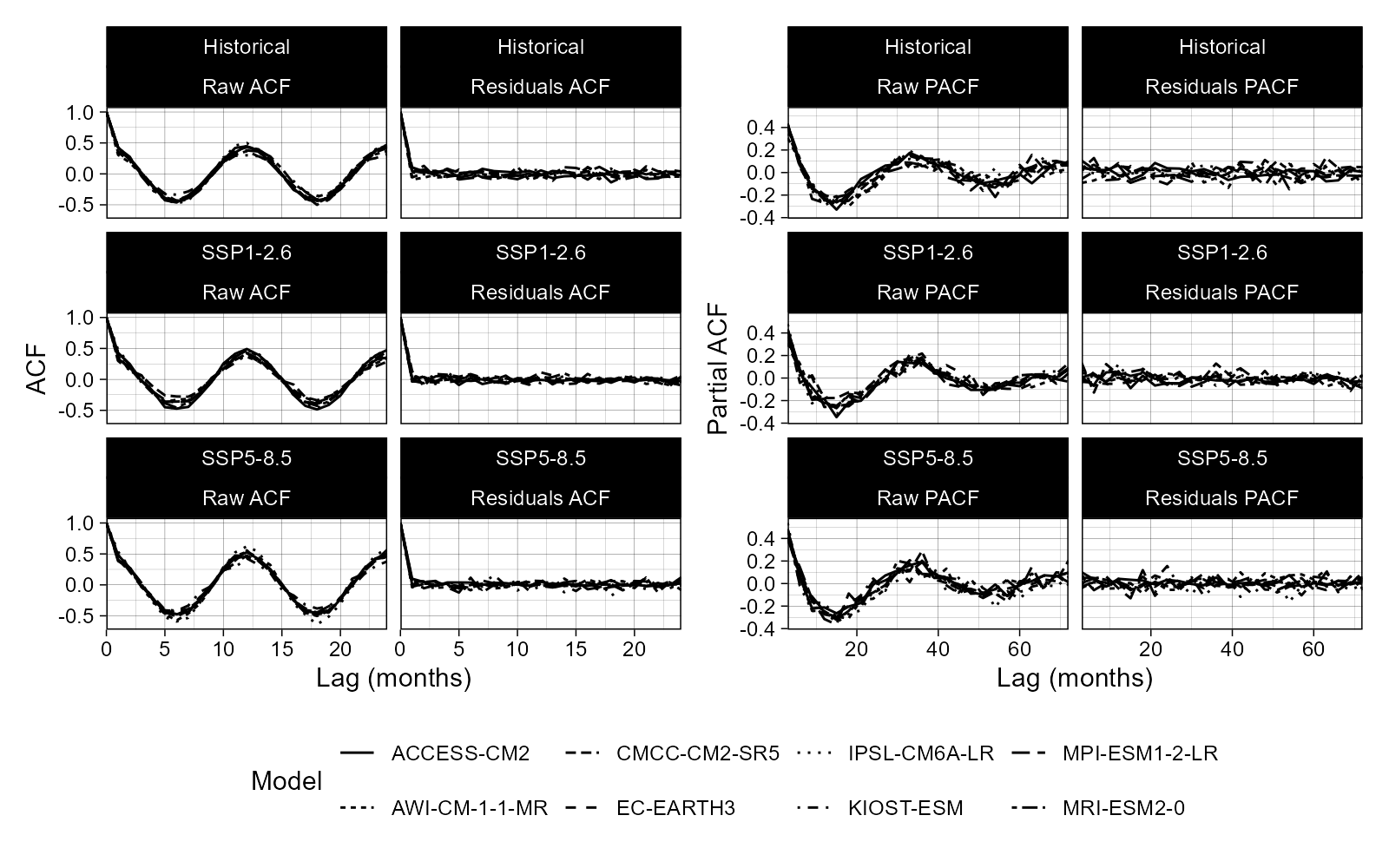}
\caption{Comparison of auto-correlation function (ACF), left, and partial auto-correlation functions (PACF), right; on East English Channel, before modeling the non-stationarity of monthly maxima (left columns of each figure) and after (right columns of each figure).}
\label{fig:acf-pacf}
\end{figure}

The N-year return levels are defined (see \cite{Coles2001}), as the value which is expected to be exceeded on average \textit{once} every N years. From Equation~\ref{non-stat}, it can bee seen that this quantity is not accessible directly because we are dealing with monthly maxima. However, the distribution of yearly maxima can be computed back, by taking the product of the monthly CDF, using:
\begin{equation}
\label{eq:retlev-monthly}
    \mathbb{P}(M_n \leq x ) = \prod_{m=1}^{12}{\mathbb{P}(M_{n,m} \leq x)},
\end{equation}
which is justified if the monthly maxima are independent. We tested this hypothesis by analyzing the auto-correlation and partial auto-correlation functions. Our analysis revealed no remaining time dependence in the residuals of the models, which strongly supports the initial hypothesis. The corresponding figures for the representative point in the Eastern English Channel are shown in Figure \ref{fig:acf-pacf}. This finding is consistent with results observed in other studies (e.g. \cite{Cheynel2025, youngman2022}).

\subsection{Long-term non-stationary models}
\label{sec:nonstatmodel}
\subsubsection{Methodology}

The 100-year return value, which is the value having an annual exceedance rate of $\frac{1}{100}$, is in widespread use in design studies (\cite{leach2025}). Such a low probability of occurrence requires to base the study on sufficient amount of data, in order to reduce the statistical uncertainty associated with the extrapolation, at least 20 to 30 years of data. However, as the climate is evolving at a quick pace (\cite{lee_ipcc_2023}), the long-term stationary assumption may no longer be valid, and extensions of such models are needed. Is this study, we incorporated a non-linear, non-parametric trend in addition to the non-parametric cycles introduced earlier in Eq.~\eqref{non-stat}. Building up from this equation, we have the following model, allowing the seasonal cycle to evolve with time:
\begin{equation} \label{full-non-stat}
    M_{\text{m,y}} \sim \text{GEV}(\mu = f_1(\text{m,y}),\sigma = f_2(\text{m,y}),\xi),
\end{equation} where \textit{m} is the month and \textit{y} is the year. We will assume here that for $i = 1, 2,\ f_i$ is a spline function on the tensor space of month and year (\cite{Wood2006}), allowing to take into account both the seasonality and the change thereof as the time varies, due to the influence of a non-stationary climate. This has made possible using the $\{\text{evgam}\}$ package (\cite{youngman2022}) of the \textbf{R} programming language (\cite{baseR}).

\subsubsection{Computation of design sea-state in a non-stationary climate}

We cannot assume, using model \eqref{full-non-stat}, that the monthly maxima are identically distributed, once the month is accounted for, contrary to the derivations made in Section~\ref{retlev-monthly}. To define an equivalent return level in a changing climate, we decided to take into account the whole lifetime of the structure.

Say, without loss of generality, that the offshore wind farm is intended to stay for $D_L = 30$ years, and is design to have an annual failure rate of $P_{\text{annual}} = \frac{1}{100}$. Then, the overall failure probability over the duration $D_L$ is, assuming that the failure rate if independent of years:
\begin{equation*}
    P_{\text{survive over } D_L \text{ years} }= (1 - P_{\text{annual}})^{D_L} \approx 0.74.
\end{equation*}

In a non-stationary climate, the annual probability of default can no longer be considered constant, as it is influenced either by the internal variability of the Earth system or by the long-term climate response to anthropogenic forcing. In that case, we can write the distribution of the maxima over the whole lifetime of the structure, $D_L$ years:

\begin{equation} \label{ret-lev-non-stat}
    P(M_{D_L} \leq x) =  \prod_{y=1}^{D_L}\prod_{m=1}^{12}{\mathbb{P}(M_{m,y} \leq x)},
\end{equation}

assuming only that the maximal values across years are independent. Building on \eqref{ret-lev-non-stat}, we propose to define an \textit{equivalent design condition}, obtained by taking the appropriate quantile of the distribution of the maximal value over the lifetime of the structure to obtain the same default probability over the lifetime, in a stationary climate, e.g. $\approx$ 0.74 in the example above.

\section{Results}

We now present the results obtained from fitting the non-stationary models to the wave data. For clarity, we report detailed findings only for a representative location in the Eastern English Channel; results for all other sites are provided in the online supplementary material.

\begin{figure}[ht]
\includegraphics[width=\textwidth]{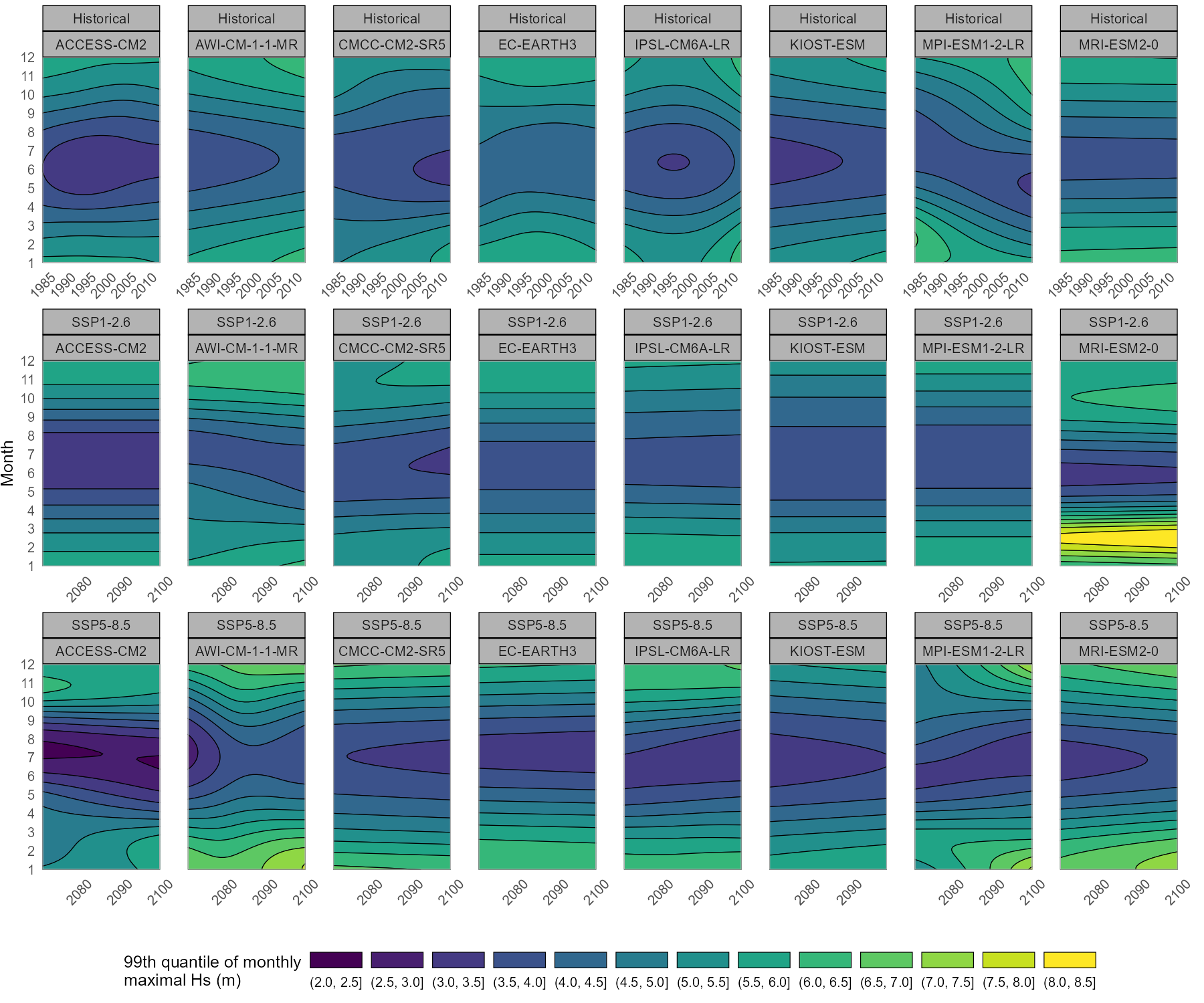}
\caption{Comparison of monthly 99th quantile, at the East English Channel Representative point, obtained from the completely non-stationary model of eq.~\ref{full-non-stat}.}
\label{fig:monthly-Q99-SRP1}
\end{figure}

Figure~\ref{fig:monthly-Q99-SRP1} displays the $99^{\text{th}}$ quantile, calculated using eq.~\ref{full-non-stat}, and estimated for each GCM at the East English Channel location. As stated in section~\ref{sec:nonstatmodel}, this quantity is dependent on the month and on the year, as a function of both the position and scale parameters. The years are represented on the x-axis, while the y-axis contains the months, from January (bottom of each subplot), to December. It is important to notice in this figure that the time period is discontinuous between the historical and future period.

The figure clearly illustrates the pronounced seasonality of significant wave height ($H_s$), with consistently higher values during winter (yellow) compared to summer (blue) across all models and time periods. A seasonal contrast in $H_s$ emerges between the historical period and scenario SSP1-2.6, and becomes even more pronounced under SSP5-8.5. Historical $H_s$ values generally range from 3 to 7 m, with gradual transitions represented by thick bands in Figure 4. Under SSP1-2.6, the range remains similar (3 m to 7 m), but seasonal variability increases, as indicated by narrower bands. In contrast, SSP5-8.5 exhibits a wider range (2 m to 8 m) and sharper seasonal fluctuations, with $H_s$ shifting from about 2.5 m in August to 5 m in September, for example. These results suggest a trend towards more intense seasonal contrasts, characterized by higher winter values (up to 8 m) and lower summer values (around 2 m), particularly evident when comparing the historical period to SSP5-8.5.

Furthermore, the figure reveals a time-evolution of the seasonality: for most models, the summer period tends to become longer and to peak later in the year, though exceptions exist (AWI, KIOST, and MRI models). Finally, the models successfully reduce to the stationary case when the data shows no evidence of shifting seasonality (e.g., KIOST model under SSP1-2.6).

\begin{figure}[ht]
\includegraphics[width=\textwidth]{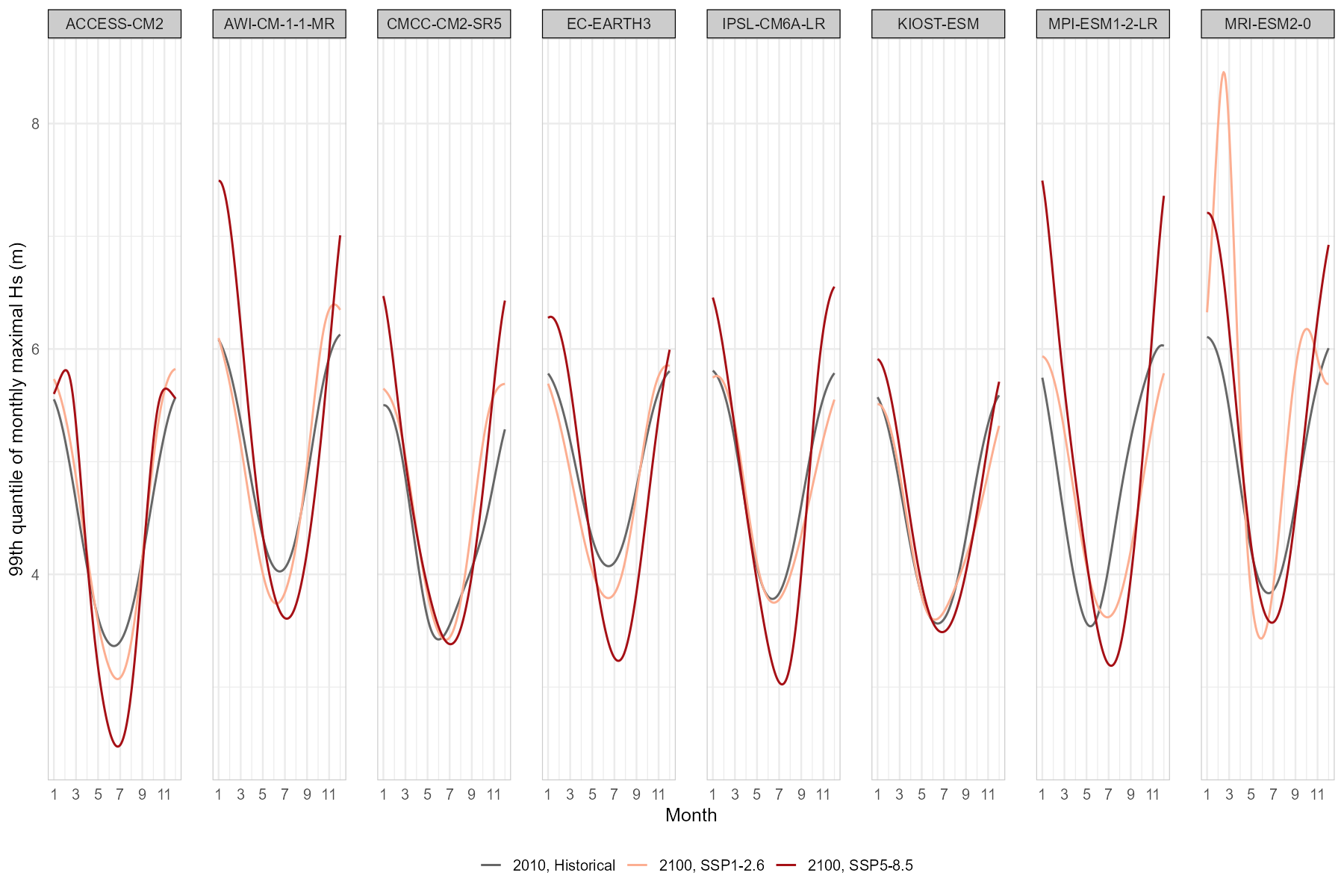}
\caption{Comparison of monthly 99th quantiles, at the East English Channel Representative point, obtained from the completely non-stationary model of eq.~\ref{full-non-stat} at different years (2010 for Historical run, 2100 for Scenarios SP1-2.6 and SSP5-8.5).}
\label{fig:compar-monthly-Q99-SRP1}
\end{figure}

\NR{An additional remark can be made regarding the results of MRI-ESM2-0 under scenario SSP1-2.6, for which the seasonality is very pronounced in the projection period, with a very large shift in the seasonal cycle (see Fig.~\ref{fig:compar-monthly-Q99-SRP1}). This behaviour is also observed at the other locations (see the online supplementary material). As this behaviour is unexpected and cannot be explained by any physical mechanism, this model should perhaps be disregarded when drawing a precise assessment of the evolution of extreme sea states under climate change. However, since the focus of this paper is on the methodology, we decided to retain this model in the remainder of the analysis.}


Figure~\ref{fig:compar-monthly-Q99-SRP1} illustrates the shift in seasonality discussed earlier by showing, for each model, the monthly extreme quantiles estimated with the non-stationary approach for two reference years: 2010, corresponding to the end of the historical period, and 2100, representing the end of the future period. The most notable feature highlighted by this figure is the seasonal shift observed in most models when comparing the grey curve (2010) with the colored curves (2100 under two scenarios). The latter exhibit lower values during summer and higher values in winter. Another striking aspect is the asymmetry between the beginning and end of the year in future scenarios, with January values exceeding December by approximately 0.2 to 0.5~m, depending on the model. Furthermore, the minimum $H_s$ values tend to occur later in the year, and the adjacent slopes become steeper, indicating a more pronounced and extended summer season compared to the historical period. Overall, all models suggest a shift of the seasonal cycle toward the end of the year, implying a potential redefinition of conventional seasons. For instance, the lowest $H_s$ values now occur in July rather than June. Finally, the comparison between SSP1-2.6 and SSP5-8.5 reveals that the latter amplifies these effects. While all models generally agree on this trend, the KIOST model shows the smallest changes, whereas the MPI model exhibits the largest.


\begin{figure}[ht]
\includegraphics[width=\textwidth]{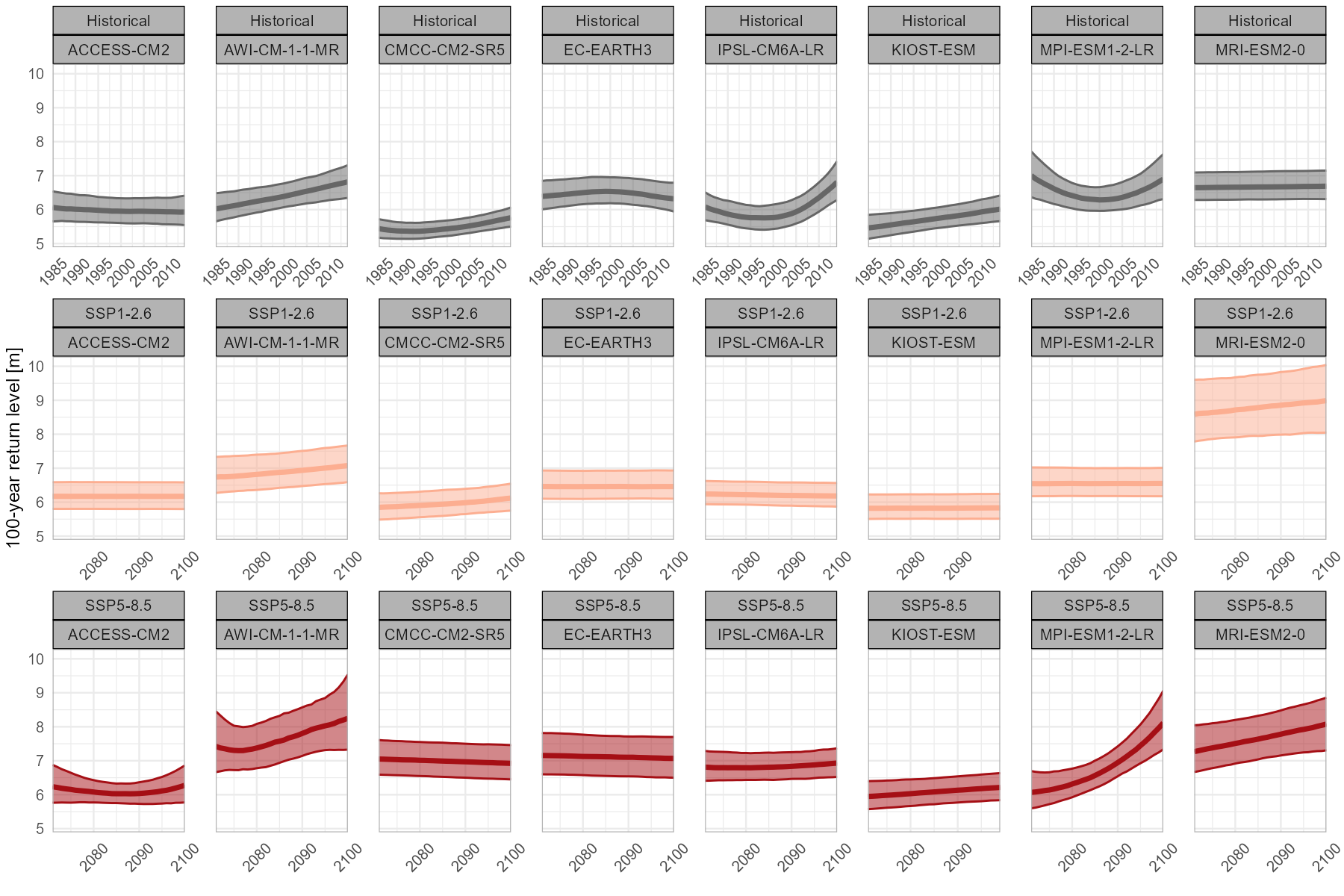}
\caption{Comparison of yearly return levels, at the East English Channel Representative point}
\label{fig:yearly-RL-SRP1}
\end{figure}

Because of the shifts identified both in the summer and winter values, it may be difficult to assess if the annual return levels also evolve with time, because the two effects can compensate each other. This can be investigated in Figure~\ref{fig:yearly-RL-SRP1}. Here, we integrate over the season using Equation~\ref{eq:retlev-monthly} of section~\ref{retlev-monthly} before inverting this quantile function to obtain return levels, which still depends on the year. Uncertainties are estimated using a Monte-Carlo approach, by simulating from the fitted model.

We see that the shift of the summer and intensification of the winter may lead to an increase of the annual return level for AWI, MRI and also MPI but for SSP5-8.5 only. While some models as ACCESS-CM2, EC-EARTH, IPSL and KIOST can also indicate that there are no long-term evolution of the return levels. And finally, the CMCC model shows a contrasted result with an increase of the return level under SSP1-2.6 and a decrease under SPP5-8.5.

We can also notice a more complex pattern in the MPI model during the historical period, which would not have been captured by a classical linear approach. Looking back at Figure~\ref{fig:monthly-Q99-SRP1}, this behavior results from a shift of the summer season towards the beginning of the year, followed by an intensified winter that fails to compensate, producing the characteristic "inverted bell curve'.

\begin{figure}[ht]
\includegraphics[width=\textwidth]{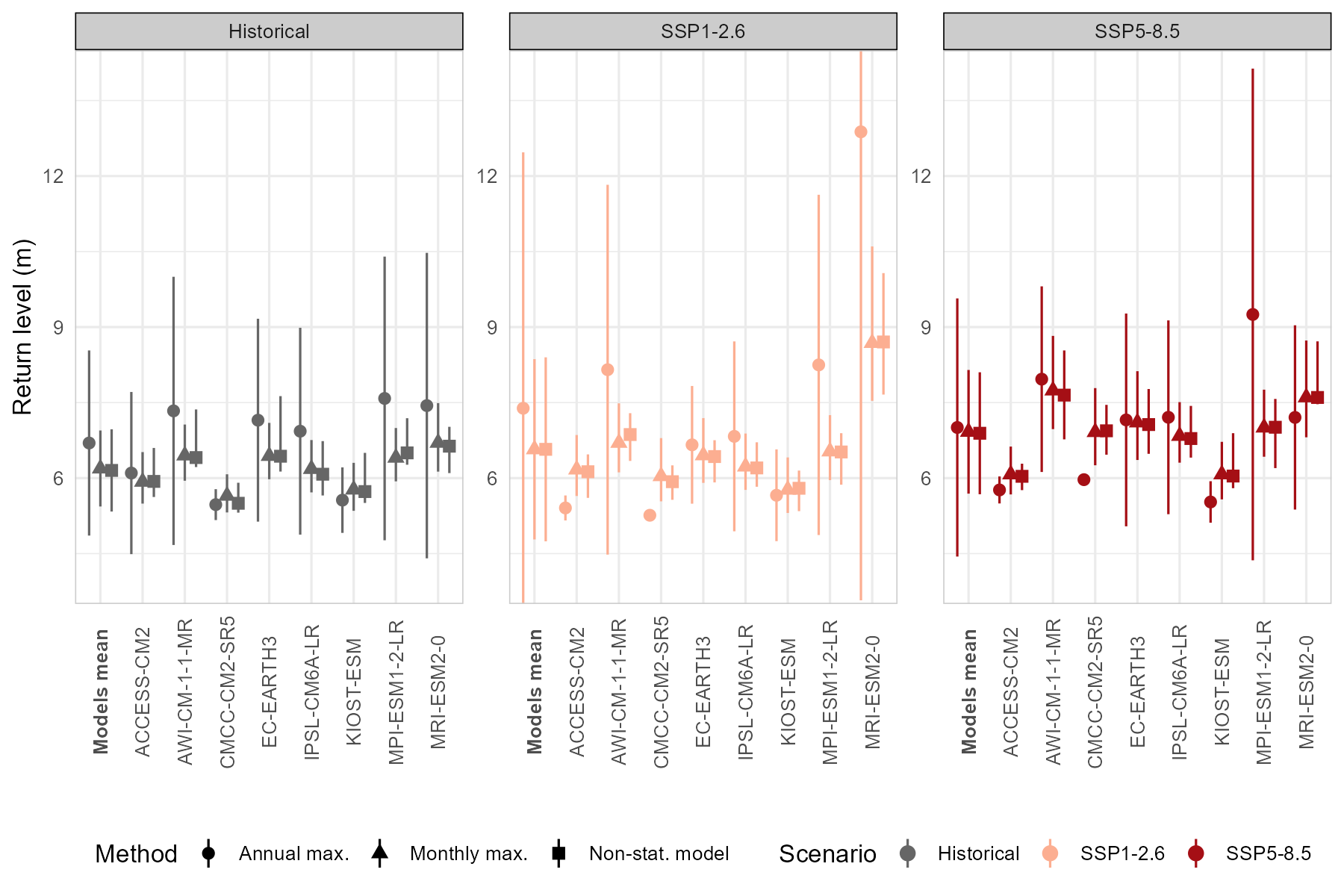}
\caption{Comparison of yearly return levels, at the East English Channel Representative point}
\label{fig:comparaisons-RL-SRP1}
\end{figure}

In order to compare the estimated values to the ones obtained using a classical stationary model, we estimated the \textit{equivalent design condition} using Equation~\eqref{ret-lev-non-stat}. In a stationary climate, this definition would coincide with the 100-year return level. These quantities are compared on Figure~\ref{fig:comparaisons-RL-SRP1}, for the different models and scenarios. We included the 100-year return level estimated from annual maxima (points), monthly maxima (triangles) and the proposed non-stationary approach (squares).

The main insight from this figure is that using monthly maxima instead of annual maxima significantly reduces the uncertainty in estimating 100-year return levels. This improvement is largely due to the model being fitted to twelve times more data, despite its increased complexity. The figure also shows strong consistency across methods, with estimates generally falling within each other's confidence intervals. In contrast, the annual maxima approach may yield spurious results because of the limited data available for fitting the GEV model in this case.

For stationary cases as identified before (e.g. KIOST model under scenario SSP1-2.6), all three approaches produce similar estimates, with the fully non-stationary model providing the narrowest confidence intervals. When averaging return levels across all models, we obtain approximately 6.15 m for the historical period, 6.75 m for SSP1-2.6, and 7 m for SSP5-8.5, indicating that future scenarios predict higher return levels.

\begin{figure}[ht]
\includegraphics[width=\textwidth]{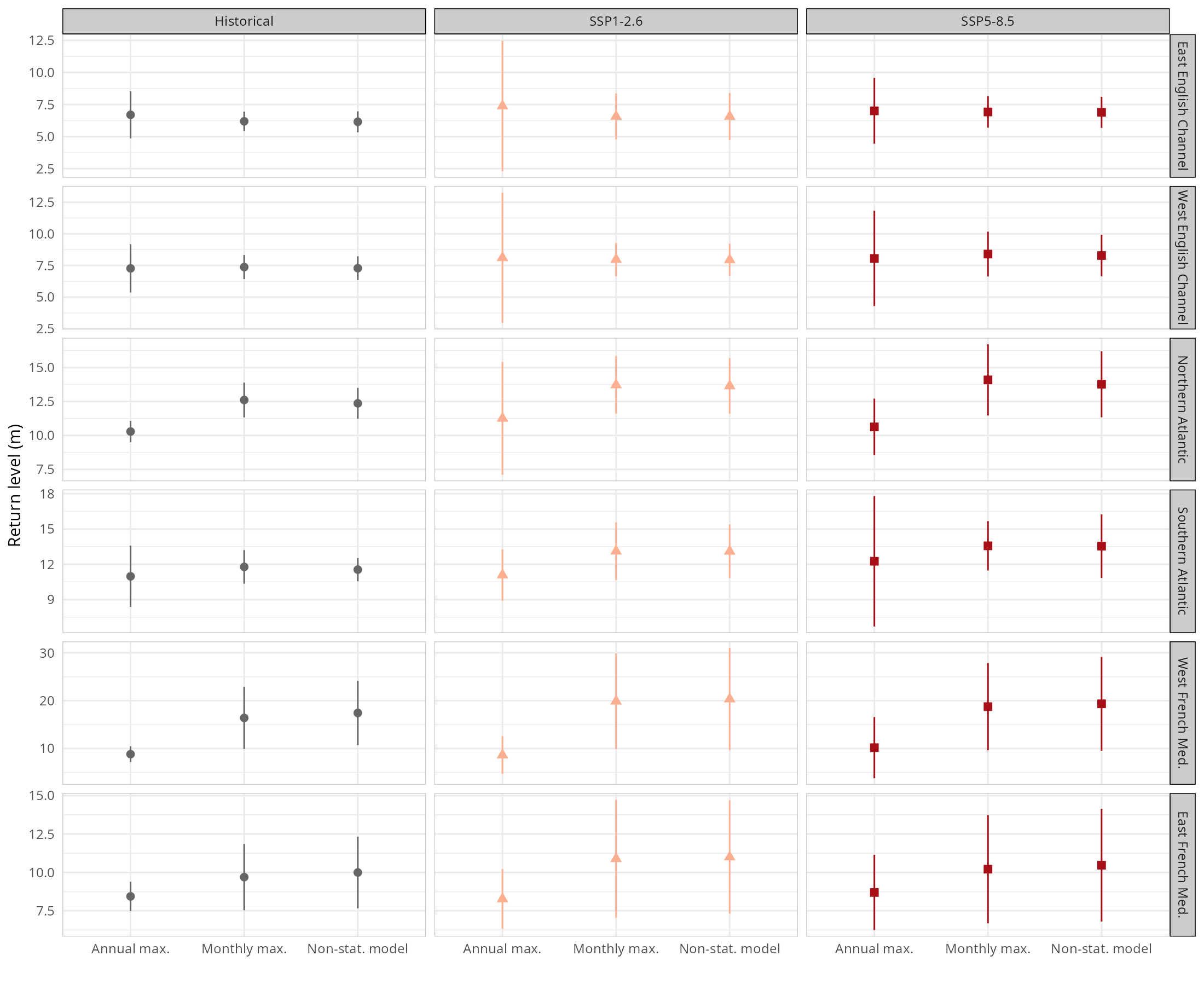}
\caption{Comparison of average 100-yr return level over the GCMs, at the different representative points and the different scenarios, with the different methods proposed.}
\label{fig:compar_RL}
\end{figure}

Lastly, Figure~\ref{fig:compar_RL} contains the estimation of the \textit{equivalent design condition} for each seafront representative point detailed previously (table \ref{tab:seafront_coords}), and using the methods used throughout the paper. All trends shown are progressive, first between the historical period and SSP1-2.6, then with SSP5-8.5, even though the return periods are very similar for the future scenarios. Only the East Channel location shows a slight decrease, the other representative points tending toward an increase. West Channel location has a slight increase, North Atlantic location has a clear increase of approximately 2.5 m, South Atlantic location has an increase of approximately 1 m, West Mediterranean location has a significant increase of 5 m for the historical scenario and 10 m for future scenarios, and finally East Mediterranean location has an increase of 1 m for the historical scenario and SSP5-8.5 and 2 m for SSP1-2.6. Globally, we can derive the same comments as the previous figure for points located in the English Chanel and in the Atlantic. The non-stationary monthly method reduces uncertainties and has values close to those of the monthly maxima method. However, for representative points in the Mediterranean Sea, the annual maxima method gives the lowest margin of uncertainty. It should be noted that West Mediterranean location has very high $H_s$ values compared to all other representative points, reaching up to 20 m and uncertainties of up to 30 m.  For the locations in the Mediterranean sea, the results are more mitigated compared to the English Channel and the Atlantic, which can be explained by a less important seasonality (\NR{see Figures~9 and 10 in the online supplementary document)} and to a less-convincing  representation of the sea-states from the GCMs in this area. In fact, the coastal processes are very important and not well represented in the very large-scale climate models used to force the numerical wave model. This has a particular impact on the Mediterranean Sea, which is surrounded by coastlines, compared to the Atlantic Ocean or the English Channel.

\begin{figure}[ht]
\includegraphics[width=\textwidth]{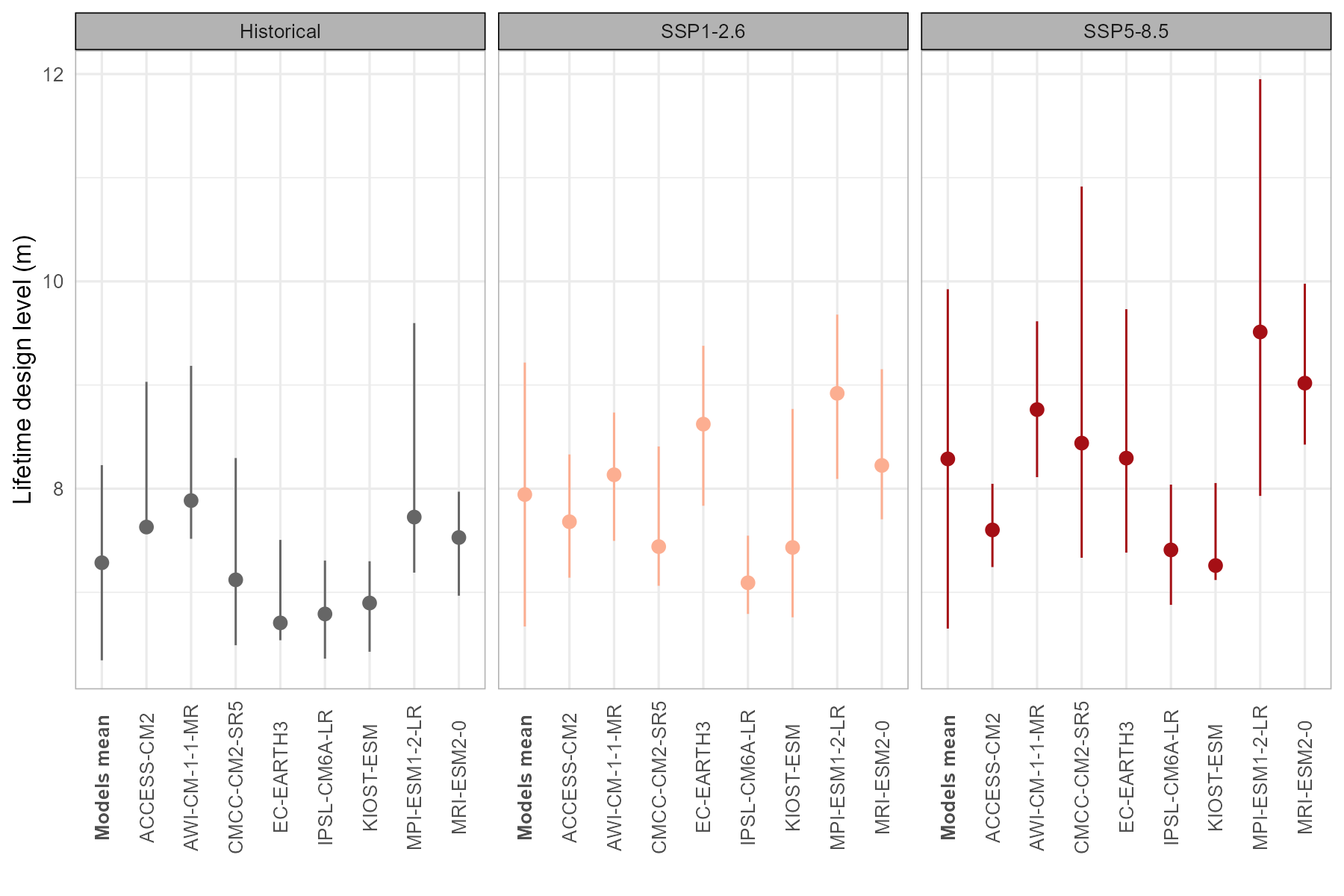}
\caption{Comparison of lifetime design level at West English Channel location, over the GCMs and the different scenarios, estimated from the fully non-stationary model.}
\label{fig:lifetime_RL_SRP2}
\end{figure}

To conclude, the lifetime design level represents the wave height to not exceed within the structure lifetime of the wind farms. In Figure \ref{fig:lifetime_RL_SRP2} the lifetime design level is represented for the English Channel. For all models the lifetime design level is smaller for the historical scenario from 1985-2014 (7.4 m) than for future scenarios from 2081 to 2100, the SSP1-2.6 (8 m) and SSP5-8.5 (8.2 m) are close to each other. The largest uncertainties are for SSP5-8.5. The models with the highest uncertainties are CMCC and MPI for the historical period and SSP5-8.5 and it is KIOST for SSP1-2.6. The investigation of the lifetime design level is interesting through all maritime seafronts (see Appendix - Figures 13 to 16). The Atlantic coast presents similarities with the English Channel even if values are higher with 11 to 13 m for the historical period and 13 to 15 m for both SSPs. The South part of the Atlantic has the lowest values compared to the North part. And for the Mediterranean Sea SSP1-2.6 seems higher than SSP5-8.5, showing really high uncertainties even if compared to all other seafronts. For the eastern part of the Mediterranean Sea, values are low around 10 m for the historical period and 12 m for the two SSPs, whereas for the West part values are high, around 20 m for all scenarios. The Mediterranean Sea highlights great uncertainties from 4 to 14 m so around 10 m difference between all models for SSP scenarios while uncertainties are lower in the Atlantic and the English Channel with a maximum of 3 m difference between all models. 

\section{Discussion}

The three French maritime seafronts — Atlantic, English Channel, and Mediterranean — display fundamentally different projection behaviors and uncertainty envelopes in climate-driven wave simulations, with significant implications for offshore wind development. Ensemble analyses and regional studies show that the Atlantic and Channel coasts generally produce more coherent signals across models, notably clearer changes in storm-track activity and a pronounced intensification of winter extremes, whereas the Mediterranean sea emerges as both a climate hotspot and as a basin with much larger inter-model spread and several physically implausible outliers in some model-wave ensembles (e.g., extremely large simulated $H_s$ in a few runs) which are likely linked to coarse GCM resolution, poor representation of local cyclogenesis, and unresolved coastal/fetch effects (\citet{hemer2013, morim2019, cos2022}). Several authors therefore caution that Mediterranean projections must be interpreted with care: ensemble medians often differ from individual model extremes, and values that appear unrealistically high typically arise from biased wind forcing or from insufficiently resolved dynamical drivers rather than from robust physical signals (\citet{giorgi2008, davison2024}). By contrast, the Atlantic basin frequently shows a distinct “deepening” of seasonality in extremes, a stronger winter peak and starker summer contrast in $H_s$, a signal that is reproduced more consistently across multiple modelling studies and wave projections (\citet{hemer2013, meucci2024}). Finally, CMIP6 (and earlier CMIP5) model families exhibit substantial inter-model differences in the representation of the large-scale drivers of wave climate (storm tracks, surface wind fields and seasonal variability), with specific models (e.g., KIOST-ESM, some CMCC configurations, MPI-ESM variants) producing divergent wind-forcing fingerprints and therefore divergent wave projections; this heterogeneity has been documented in model evaluation and ranking studies and argues for multi-model, bias-corrected and high-resolution downscaling strategies when deriving local extreme $H_s$ estimates (\citet{meucci2024, lorenzo2025}). In summary, while Atlantic and Channel projections provide relatively coherent seasonal signals —- particularly a clearer intensification of winter extremes —- Mediterranean projections remain more uncertain and prone to isolated, physically questionable extremes. Risk assessments in the Mediterranean sea therefore demand rigorous model selection, downscaling, and plausibility checks rather than blind reliance on raw GCM-driven wave outputs (\citet{hemer2013, cos2022, davison2024}).

The seasonal variability of extreme significant wave heights shows marked contrasts between France's three maritime seafronts, reflecting both atmospheric circulation patterns and basin morphology. Along the Atlantic coast and in the English Channel, studies by \citet{bricheno2018} and \citet{bulteau2013} highlight a clear winter dominance of extreme events, associated with intensified westerly storms and North Atlantic depressions, whereas summer remains generally calmer, confirming a strongly seasonal regime. These works suggest that winter maxima are projected to intensify consistent with our results suggesting the increase of severe winters and really low values of $H_s$ in summer. In fact, \citet{bricheno2018} anticipate an increase of annual maxima $H_s$ in the three French seafronts for mid-century and end of century with RCP4.5 and RCP8.5. Overall, all studies agree that the cold season remains dominant for extreme $H_s$, even though future projections point to a more complex seasonal cycle, with longer and later summers. \citet{chirosca2022}, using ERA5  from 2001 to 2020, found that $H_s$ 95 percentile values are increasing in the North Atlantic coast and in the South of the English channel. As \citet{vanem2015} discerned a shift towards higher extremes in a future wave climate similar to our results in the North Atlantic with RCP4.5 and RCP8.5 comparing 1970-1999 to 2071-2100. An intensification of the 100 return level is found in the Atlantic and the English Channel by \citet{bulteau2013} using the POT and GPD using numerical wave hindcast BOBWA database  from 1958 to 2002 in accordance with our results. Within this context, our results showing a tendency towards more intense winters and calmer yet delayed and prolonged summers, together with an increase in the return period of 100-year $H_s$ are consistent with these previous findings, suggesting a future regime characterized by stronger seasonal contrasts and a temporal redistribution of calm conditions towards early autumn.

The adoption of non-stationary extreme-value models has emerged as a critical advance in assessing changes in wave extremes, offering superior performance compared to stationary assumptions under evolving climate conditions. Traditional stationary models presume that the statistical characteristics of wave climate remain constant over time, an assumption increasingly invalid in the presence of long-term trends, changing storm regimes or evolving wind and fetch patterns as there are important in French seafronts. \citet{vanem2015} showed that including time-varying location and scale parameters in extreme-value modelling substantially improves the estimation of return levels (e.g., centennial $H_s$) under non-stationary forcing and reduces bias in projections. For RCP4.5, \citet{vanem2015} reported an uncertainty of 13.15 m for the 100-year return level, which is substantially larger than the uncertainty obtained with our method (maximum of 10 m). This reduction in uncertainty is likely attributable to the larger amount of data used in our analysis, as the increased sample size across all models leads to more robust extreme value estimates. \citet{turki2020} highlighted that non-stationary approaches can account for local, seasonal and directional variability in wave extremes across basins (e.g., North Atlantic, Mediterranean) where stationary assumptions fail to reproduce observed shifts in intensity, timing and frequency of events. More recently, \citet{ewans2023} emphasized that non-stationary extreme-value analyses are indispensable for climate adaptation and coastal engineering, since they quantify how the probability of rare, extreme wave events evolves under future scenarios rather than assuming historical stability. Complementing these findings, earlier studies (\citet{mendez2008, menendez2009}) illustrated how modelling non-stationarity in seasonal parameters yields more realistic return value estimates for marine extremes by capturing seasonal shifts in storminess and fetch, thereby supporting the use of non-stationary parameterization for extreme $H_s$. Altogether, these works converge on the conclusion that non-stationary models not only offer a more realistic statistical representation of changing wave climates but are also essential for deriving credible return periods and extreme-value projections in a non-equilibrium climate system. Our study on wave extremes therefore provides a new, effective method for monthly maxima with reduced bias and taking climate change into account. Using monthly maxima instead of annual maxima provides a finer temporal resolution that better captures intra-annual variability in wave extremes and their seasonal drivers. This approach allows the detection of shifts in the timing and intensity of extreme events, an information that is lost when only annual maxima are considered. Moreover, monthly maxima improve the robustness of non-stationary analyses, as they provide a larger sample size and make it possible to explicitly model seasonal covariates (e.g., storm seasonality, climate indices), leading to more accurate and physically meaningful estimates of extreme return levels. 

\section{Conclusion and perspectives}

This study introduces a non-parametric approach based on monthly maxima to estimate return levels for monthly and annual periods from 8 GCMs, explicitly accounting for climate change effects. A key contribution lies in proposing a new method for calculating design levels — equivalent to return periods — under non-stationary conditions driven by a changing climate. The consistency of our results with existing literature reinforces the credibility and the relevance of the approach.

By leveraging monthly maxima, we substantially reduce uncertainty in extreme-value estimation. The parsimonious, data-driven model captures seasonality and its potential shifts, as well as changes in extremes themselves, making it particularly suited for assessing rare events critical to offshore wind farm design.

Our findings reveal coherent signals of stronger seasonal contrasts under future climates, with intensified winters and calmer yet prolonged summers, especially in the Atlantic and English Channel. Despite significant inter-model uncertainty — most pronounced in the Mediterranean sea — the ensemble generally indicates higher extreme sea states under both SSP1-2.6 and SSP5-8.5 scenarios.

This work opens several avenues for future research: addressing biases in extreme-value estimation to improve historical consistency; refining wave models along the French coastline for better representation of local dynamics; developing parametric approaches to seasonality and its evolution; and exploring joint distributions of significant wave height ($H_s$) and peak period ($T_p$) under future climate scenarios to enhance the robustness of offshore wind design.

\codedataavailability{The data and code supporting the findings of this study are openly available on Zenodo at the following URL: \href{https://doi.org/10.5281/zenodo.17977552}{https://doi.org/10.5281/zenodo.17977552}. The repository contains the datasets and all scripts required to reproduce the figures presented in the paper. All analyses and visualizations were performed using the R statistical computing open-source software \cite{baseR}.
}

\authorcontribution{Conceptualization: NR, CP, YK and LD ; methodology: NR; data curation: NR, TC and CP; data visualization: NR and TC; software: NR and TC; writing original draft: NR, CP, YK. All authors approved the final submitted paper.} 

\competinginterests{The authors declare that they have no known competing financial interests or personal relationships that could have appeared to influence the work reported in this paper.} 

\begin{acknowledgements}
This research has been supported by France Energies Marines and its members and partners, as well as French State funding managed by the National Research Agency under the France 2030 investment plan (grant no. ANR-10-IEED-0006-34), within the \href{https://www.france-energies-marines.org/projets/2c-now)}{2C-NOW project}. 
\end{acknowledgements}

\bibliographystyle{copernicus}
\bibliography{FEM-2CNOW}

\end{document}


\section{Monthly quantiles at SRPs}

\begin{figure}[ht]
\includegraphics[width=\textwidth]{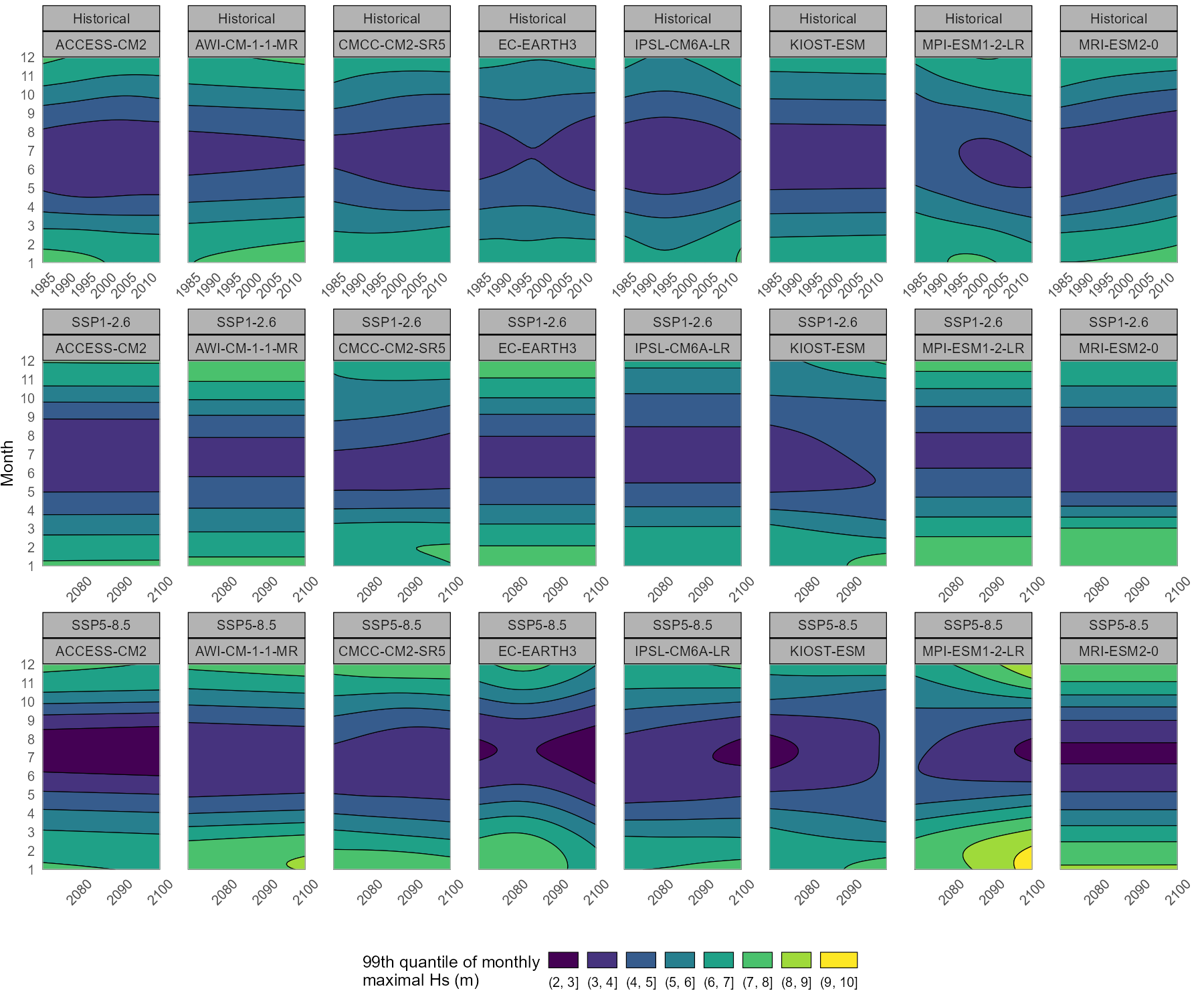}
\caption{Comparison of monthly 99th quantile, at the West English Channel Representative point}
\label{fig:monthly_RL_SRP2}
\end{figure}

\begin{figure}[ht]
\includegraphics[width=\textwidth]{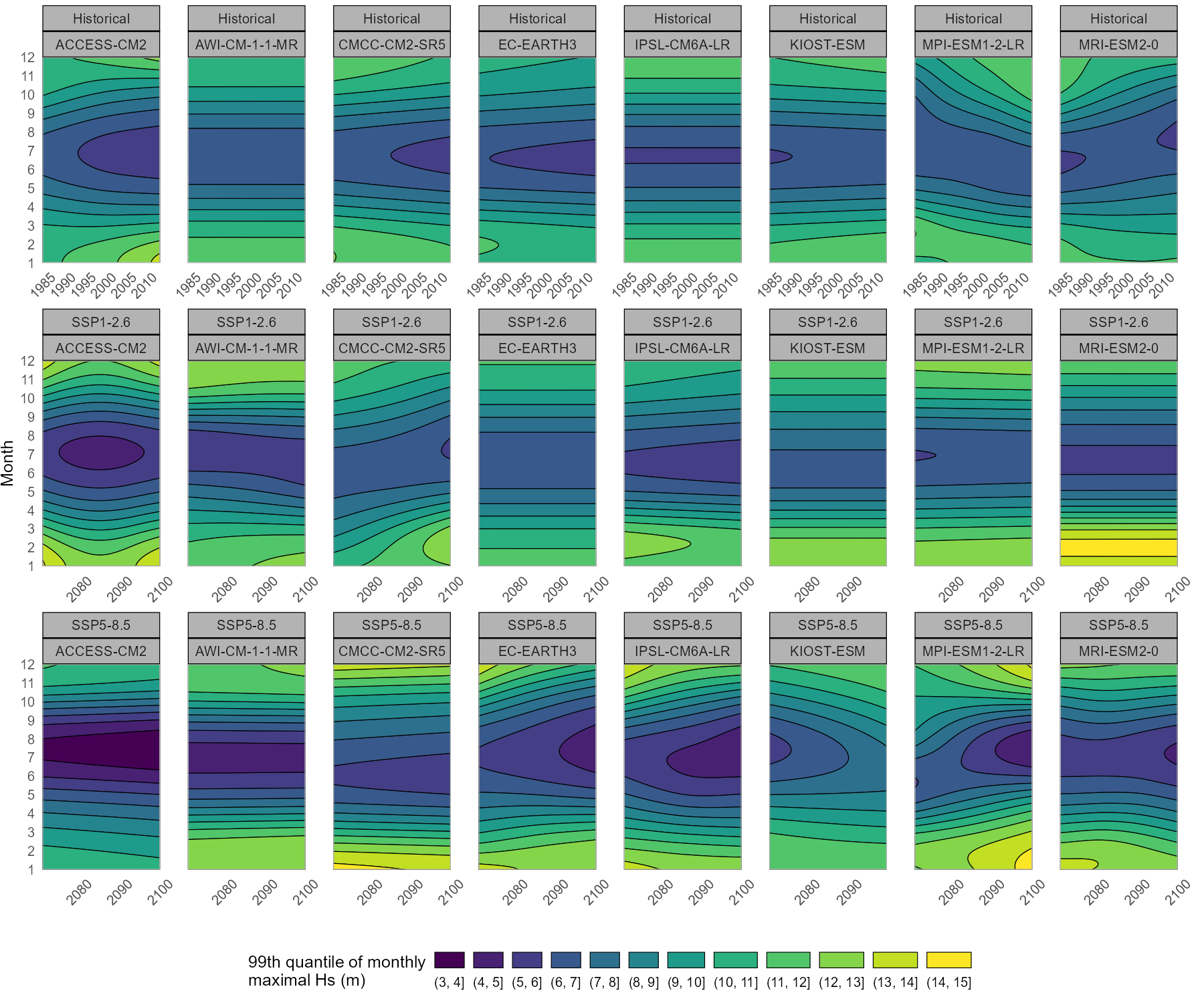}
\caption{Comparison of monthly 99th quantile, at the North Atlantic Representative point}
\label{fig:monthly_RL_SRP3}
\end{figure}

\begin{figure}[ht]
\includegraphics[width=\textwidth]{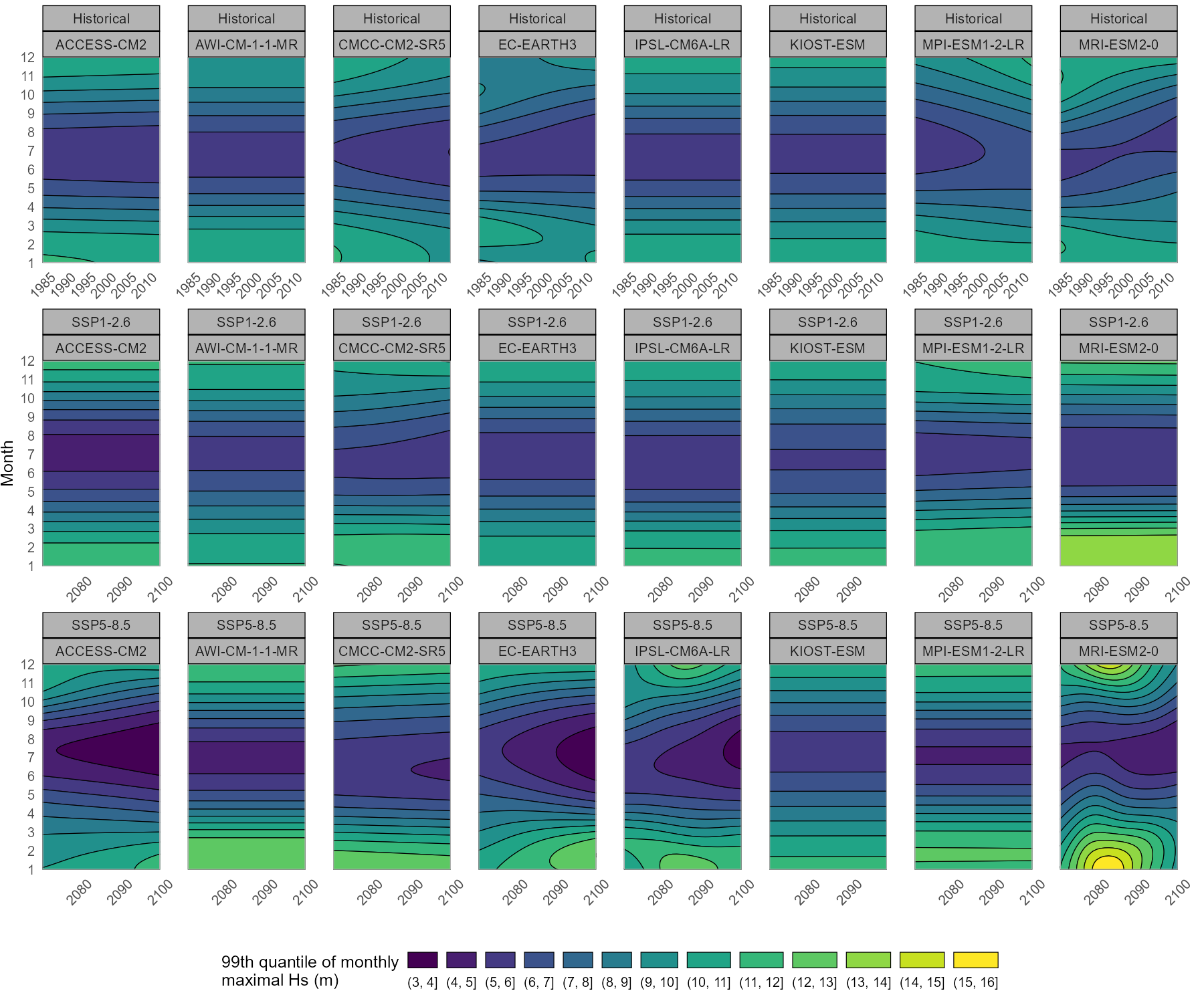}
\caption{Comparison of monthly 99th quantile, at the South Atlantic Representative point}
\label{fig:monthly_RL_SRP4}
\end{figure}

\begin{figure}[ht]
\includegraphics[width=\textwidth]{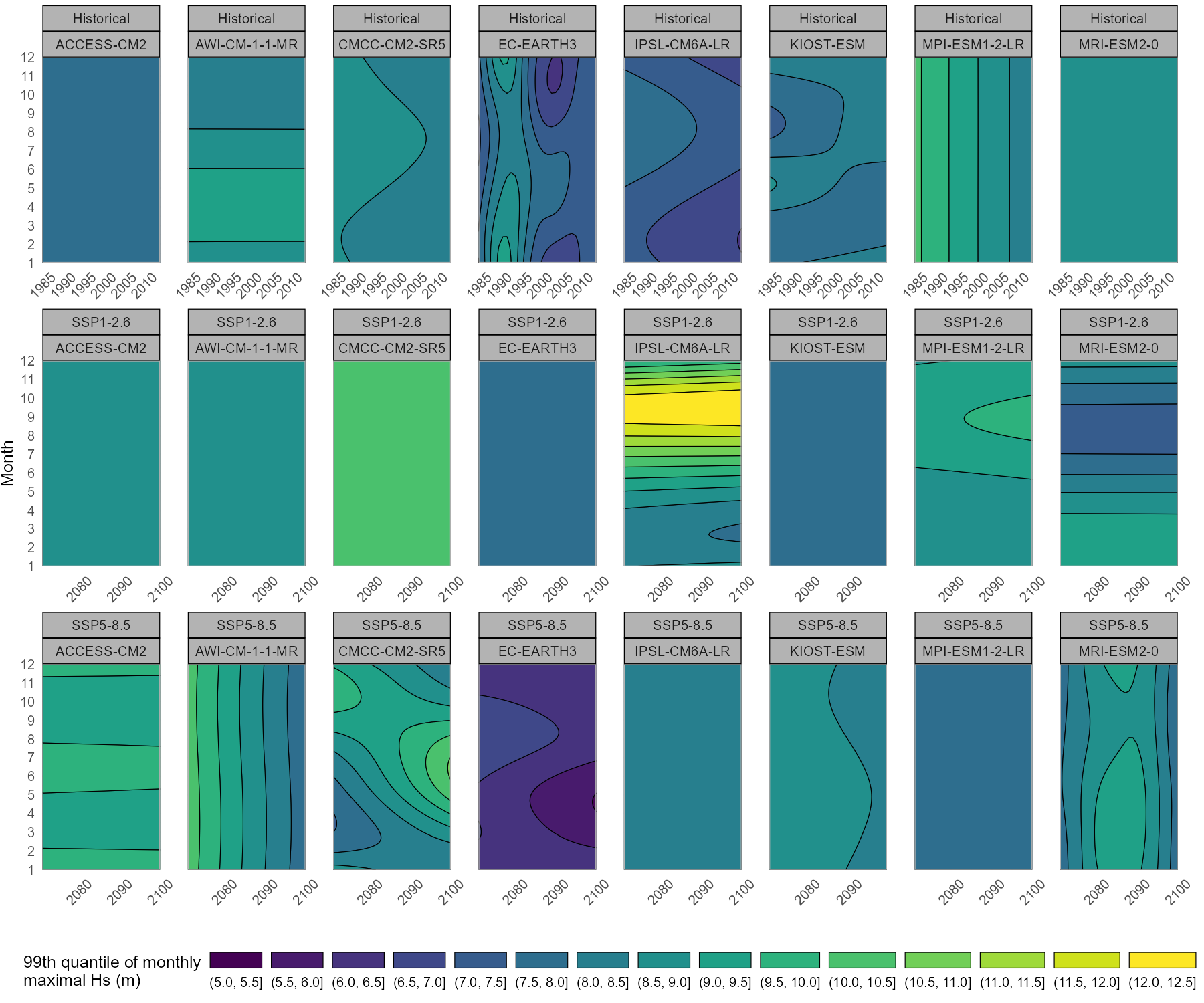}
\caption{Comparison of monthly 99th quantile, at the West Mediterranean Representative point}
\label{fig:monthly_RL_SRP5}
\end{figure}

\begin{figure}[ht]
\includegraphics[width=\textwidth]{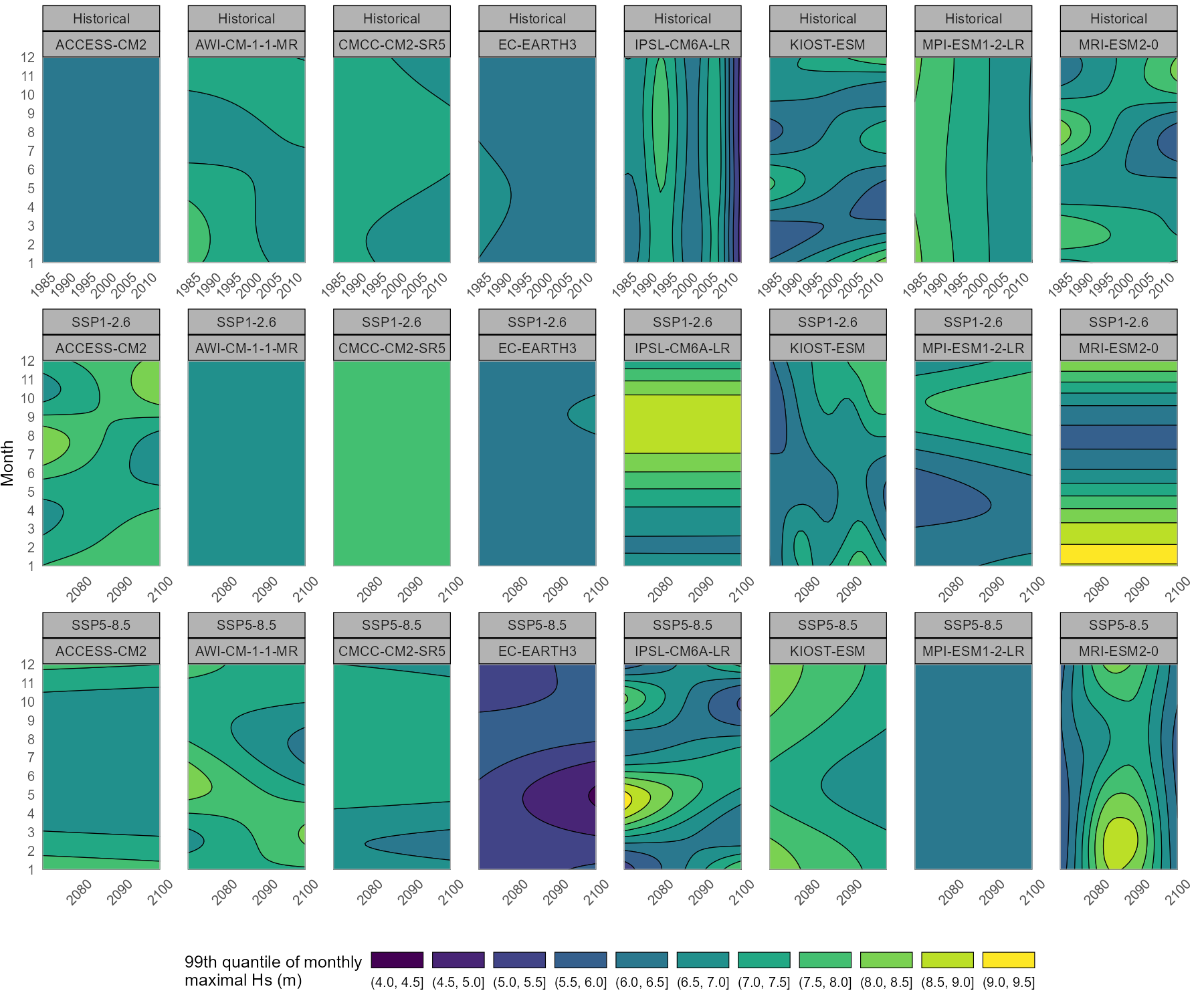}
\caption{Comparison of monthly 99th quantile, at the East Mediterranean Representative point}
\label{fig:monthly_RL_SRP6}
\end{figure}

\FloatBarrier

\section{Comparison of monthly quantiles at SRPs for given years}

\begin{figure}[ht]
\includegraphics[width=\textwidth]{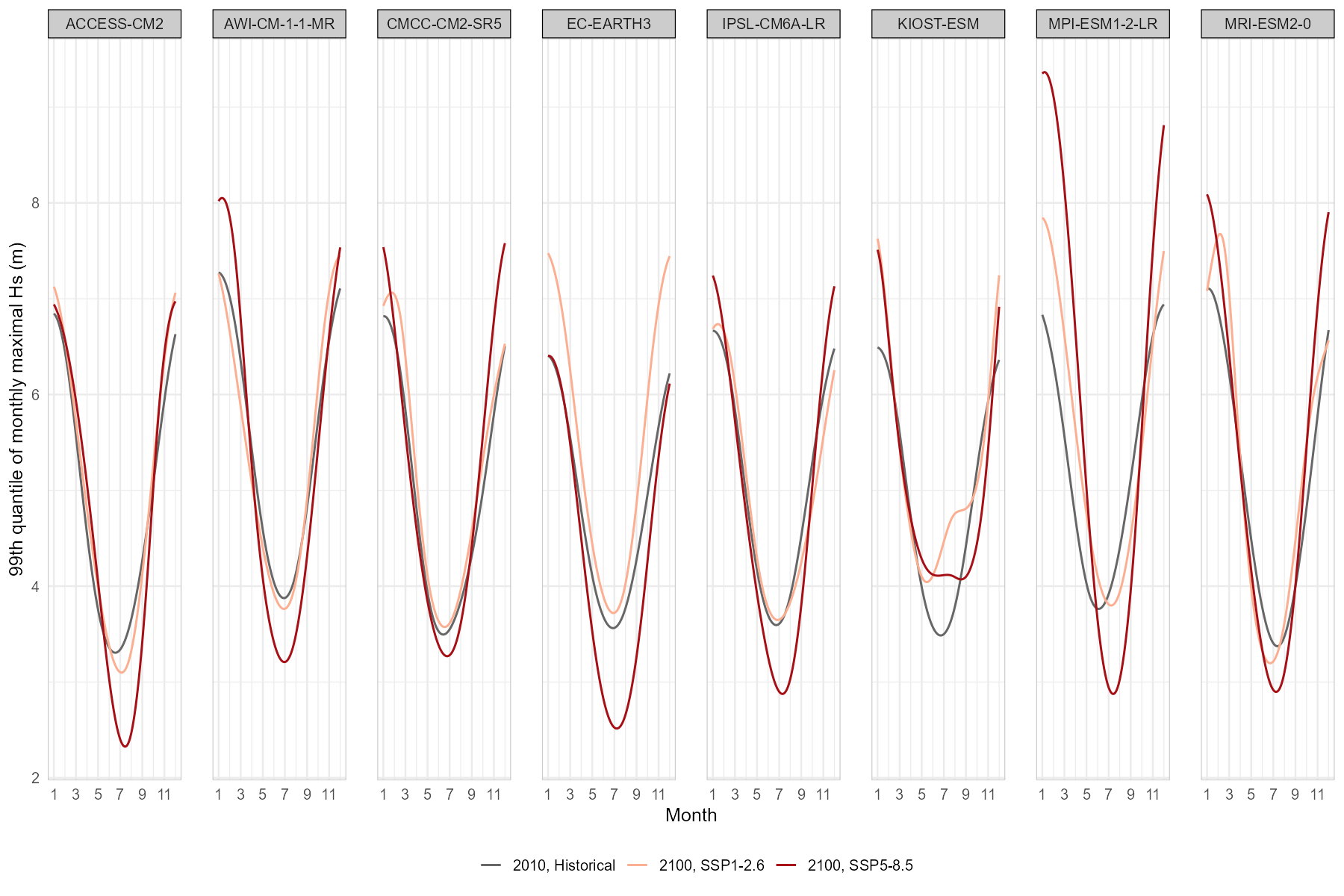}
\caption{Comparison of monthly 99th quantile, at the West English Channel Representative point, at given years (2010 for Historical run, 2100 for Scenarios SP1-2.6 and SSP5-8.5)}
\label{fig:compar_monthly_RL_SRP2}
\end{figure}

\begin{figure}[ht]
\includegraphics[width=\textwidth]{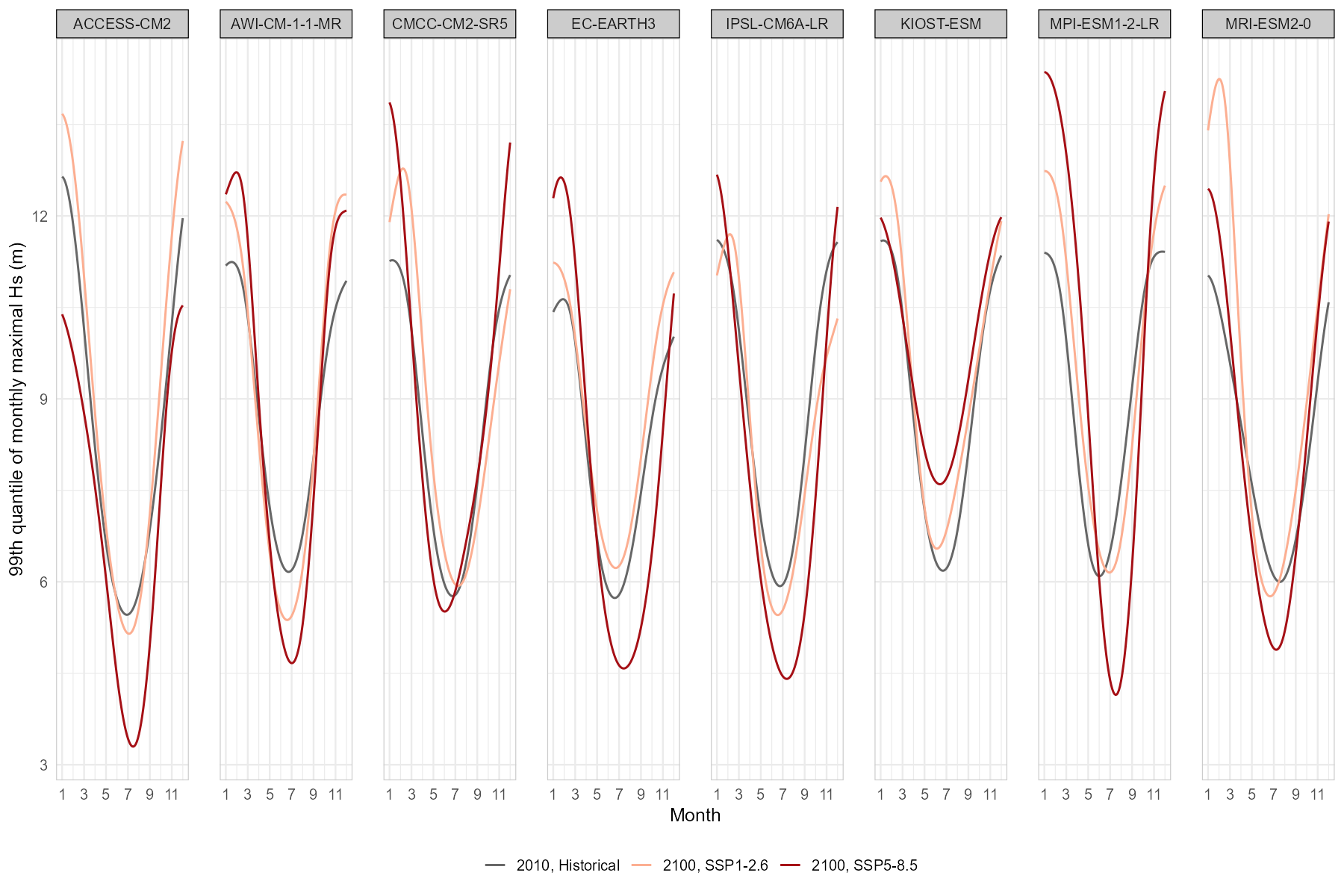}
\caption{Comparison of monthly 99th quantile, at the North Atlantic Representative point, at given years (2010 for Historical run, 2100 for Scenarios SP1-2.6 and SSP5-8.5)}
\label{fig:compar_monthly_RL_SRP3}
\end{figure}

\begin{figure}[ht]
\includegraphics[width=\textwidth]{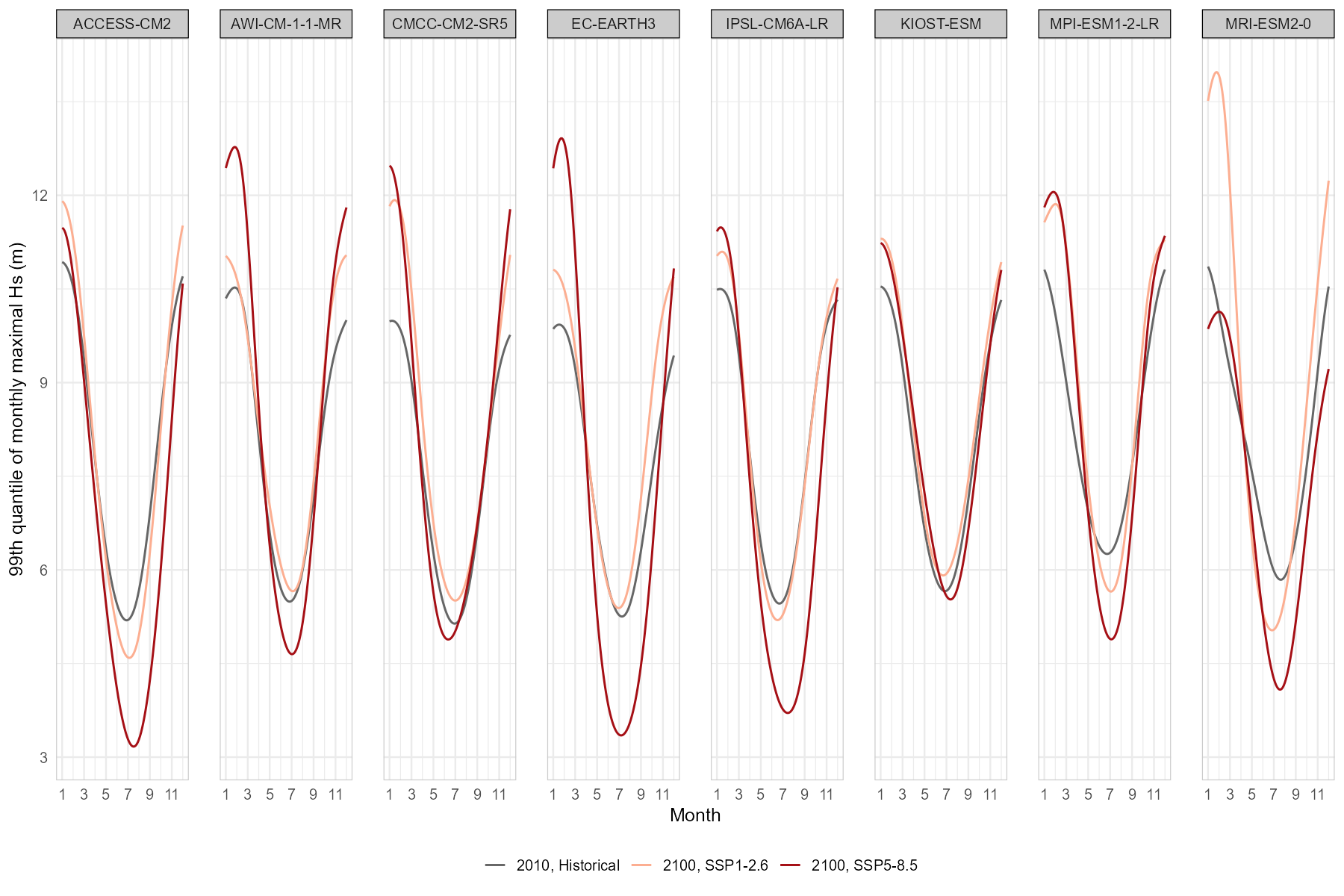}
\caption{Comparison of monthly 99th quantile, at the South Atlantic Representative point, at given years (2010 for Historical run, 2100 for Scenarios SP1-2.6 and SSP5-8.5)}
\label{fig:compar_monthly_RL_SRP4}
\end{figure}

\begin{figure}[ht]
\includegraphics[width=\textwidth]{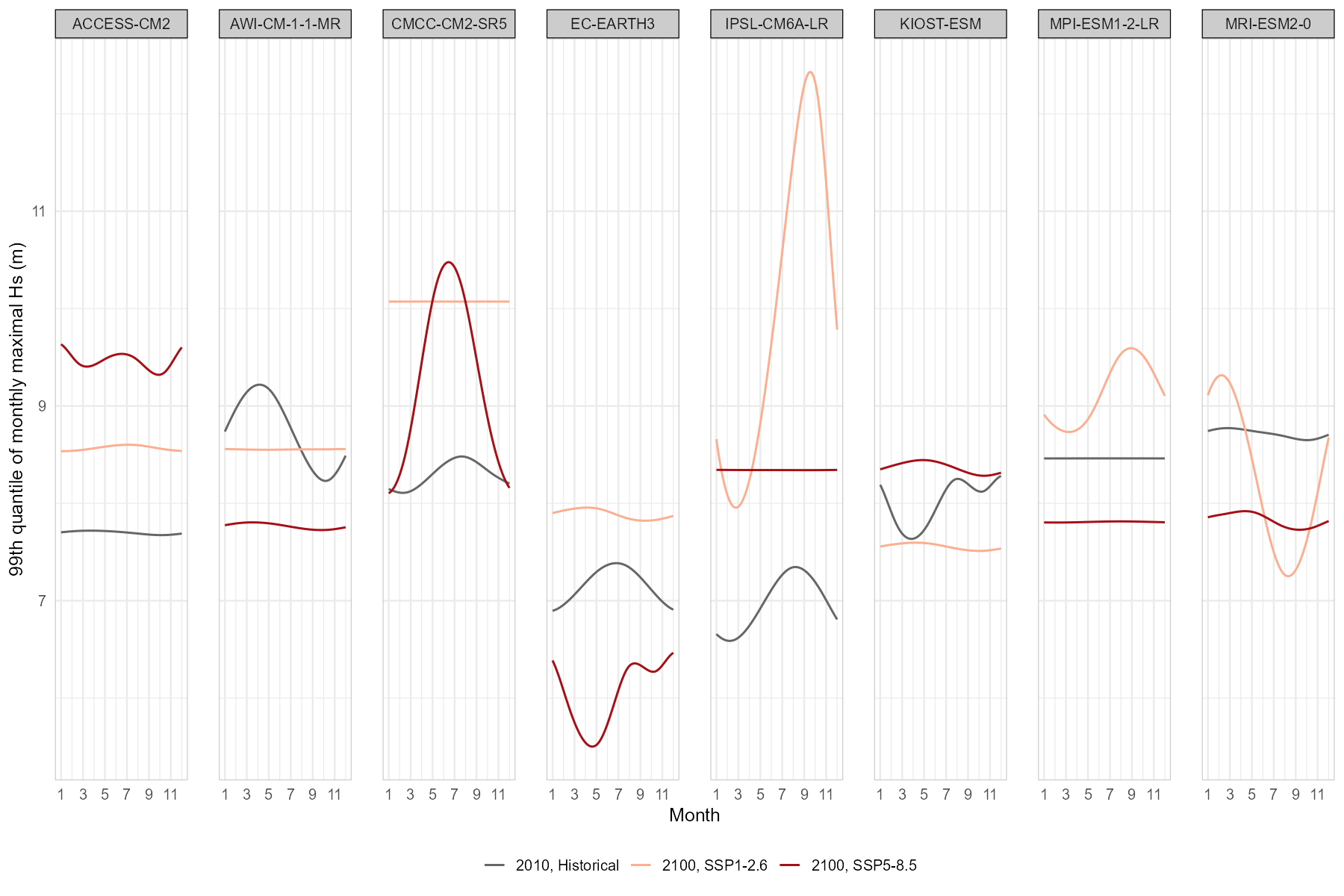}
\caption{Comparison of monthly 99th quantile, at the West Mediterranean Representative point, at given years (2010 for Historical run, 2100 for Scenarios SP1-2.6 and SSP5-8.5)}
\label{fig:compar_monthly_RL_SRP5}
\end{figure}

\begin{figure}[ht]
\includegraphics[width=\textwidth]{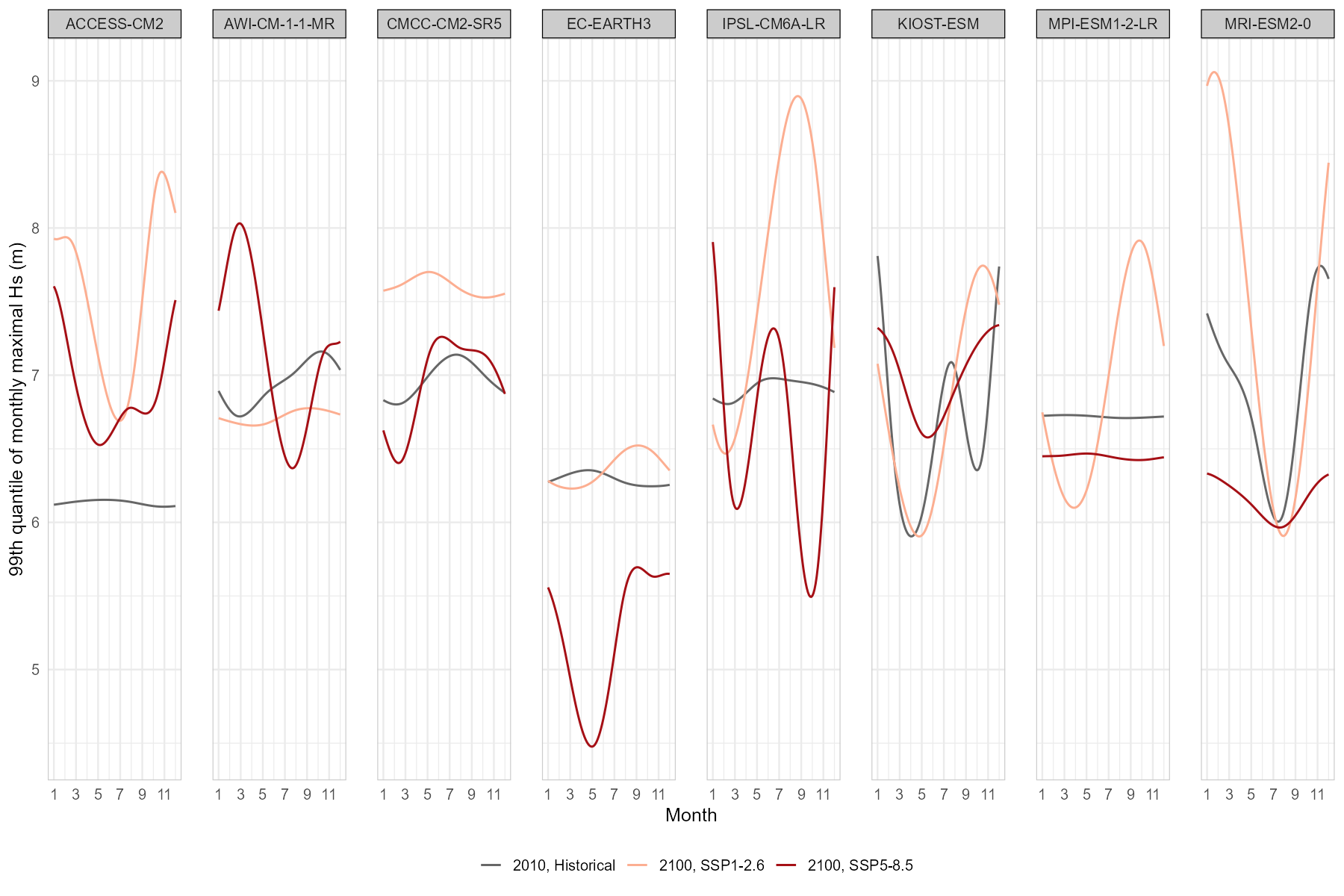}
\caption{Comparison of monthly 99th quantile, at the East Mediterranean Representative point, at given years (2010 for Historical run, 2100 for Scenarios SP1-2.6 and SSP5-8.5)}
\label{fig:compar_monthly_RL_SRP6}
\end{figure}

\FloatBarrier

\section{Yearly return levels at the different SRPs}

\begin{figure}[ht]
\includegraphics[width=\textwidth]{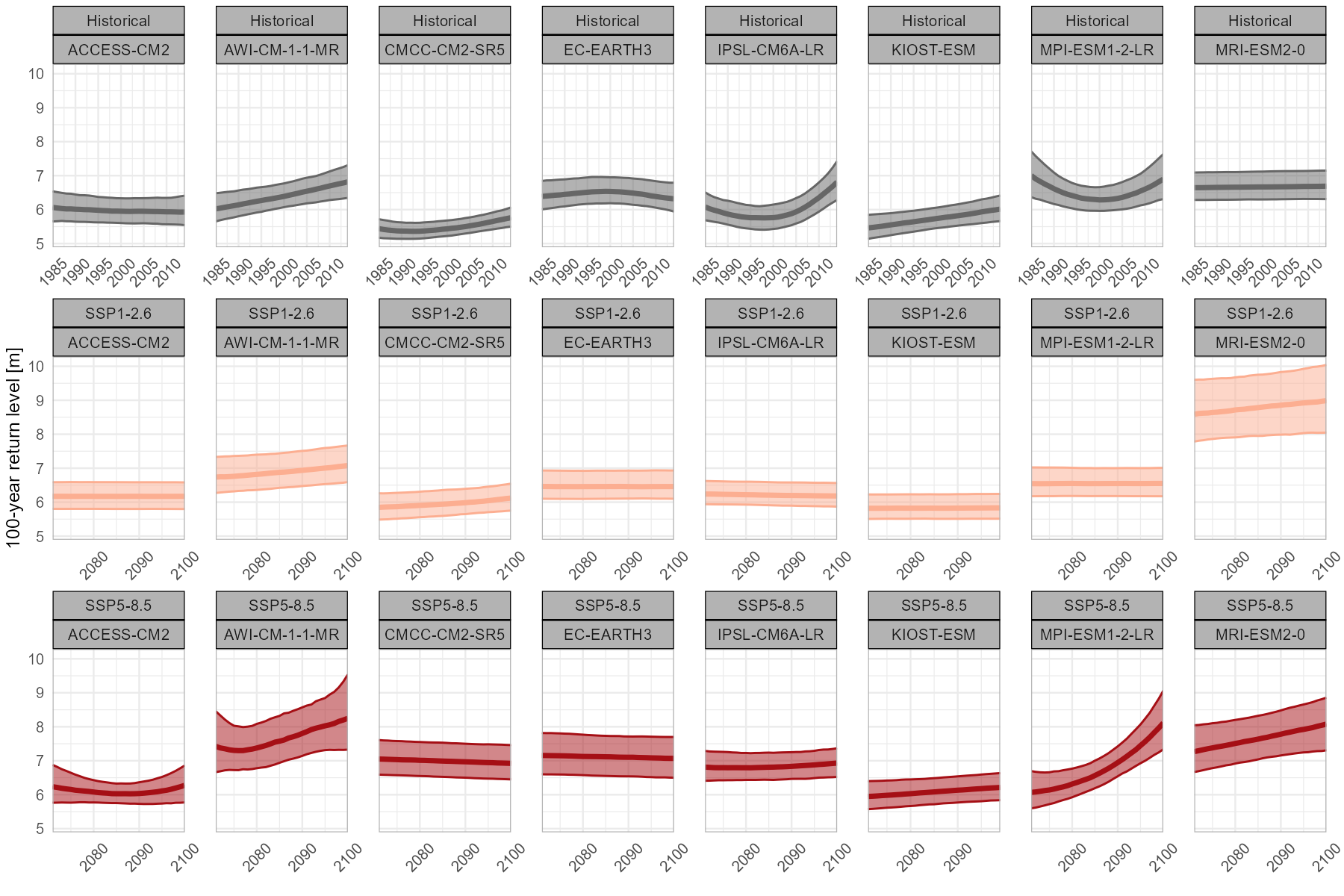}
\caption{Comparison of yearly return levels, at the East English Channel Representative point}
\label{fig:yearly_RL_SRP1}
\end{figure}

\begin{figure}[ht]
\includegraphics[width=\textwidth]{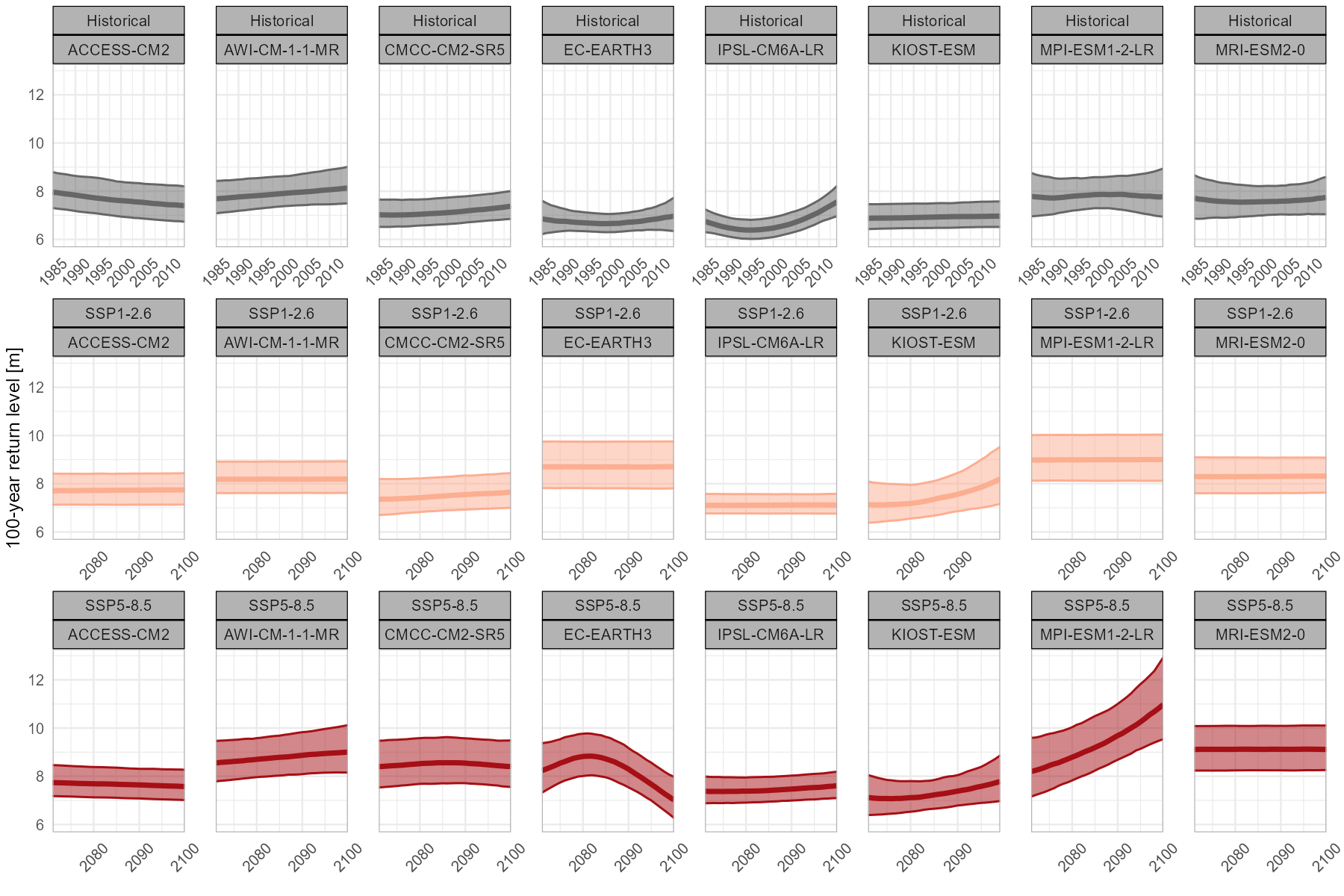}
\caption{Comparison of yearly return levels, at the West English Channel Representative point}
\label{fig:yearly_RL_SRP2}
\end{figure}

\begin{figure}[ht]
\includegraphics[width=\textwidth]{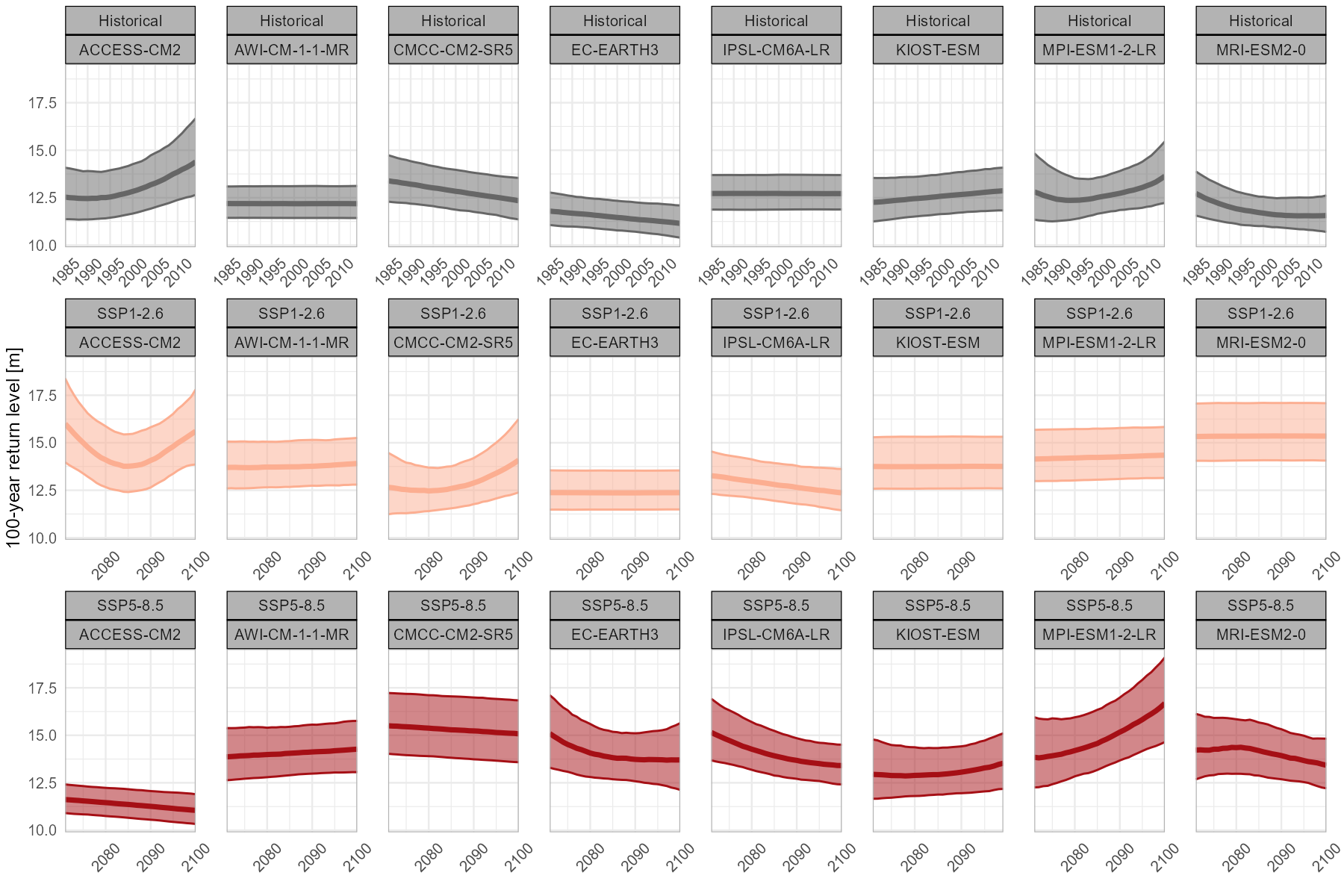}
\caption{Comparison of yearly return levels, at the North Atlantic Representative point}
\label{fig:yearly_RL_SRP3}
\end{figure}

\begin{figure}[ht]
\includegraphics[width=\textwidth]{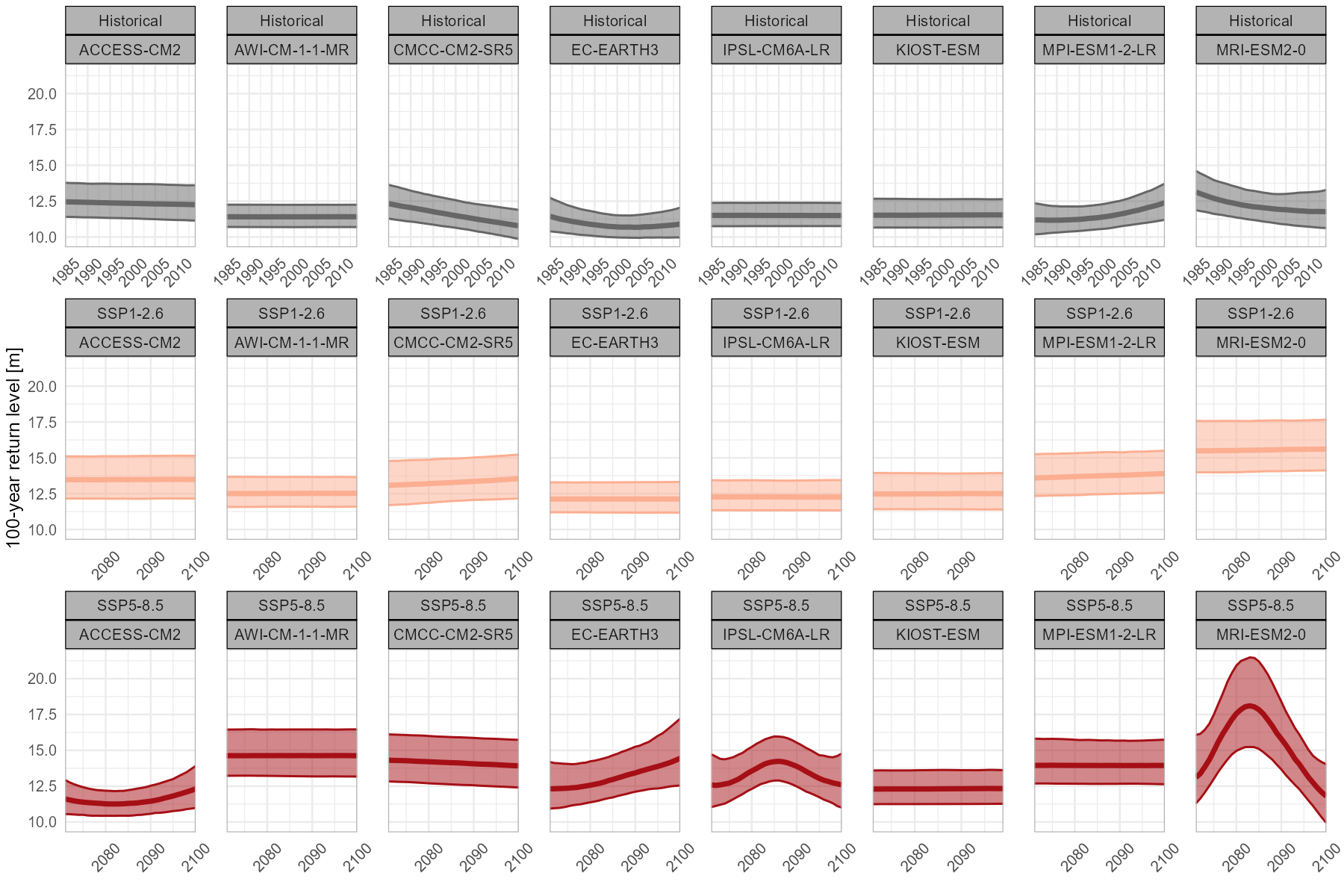}
\caption{Comparison of yearly return levels, at the South Atlantic Representative point}
\label{fig:yearly_RL_SRP4}
\end{figure}

\begin{figure}[ht]
\includegraphics[width=\textwidth]{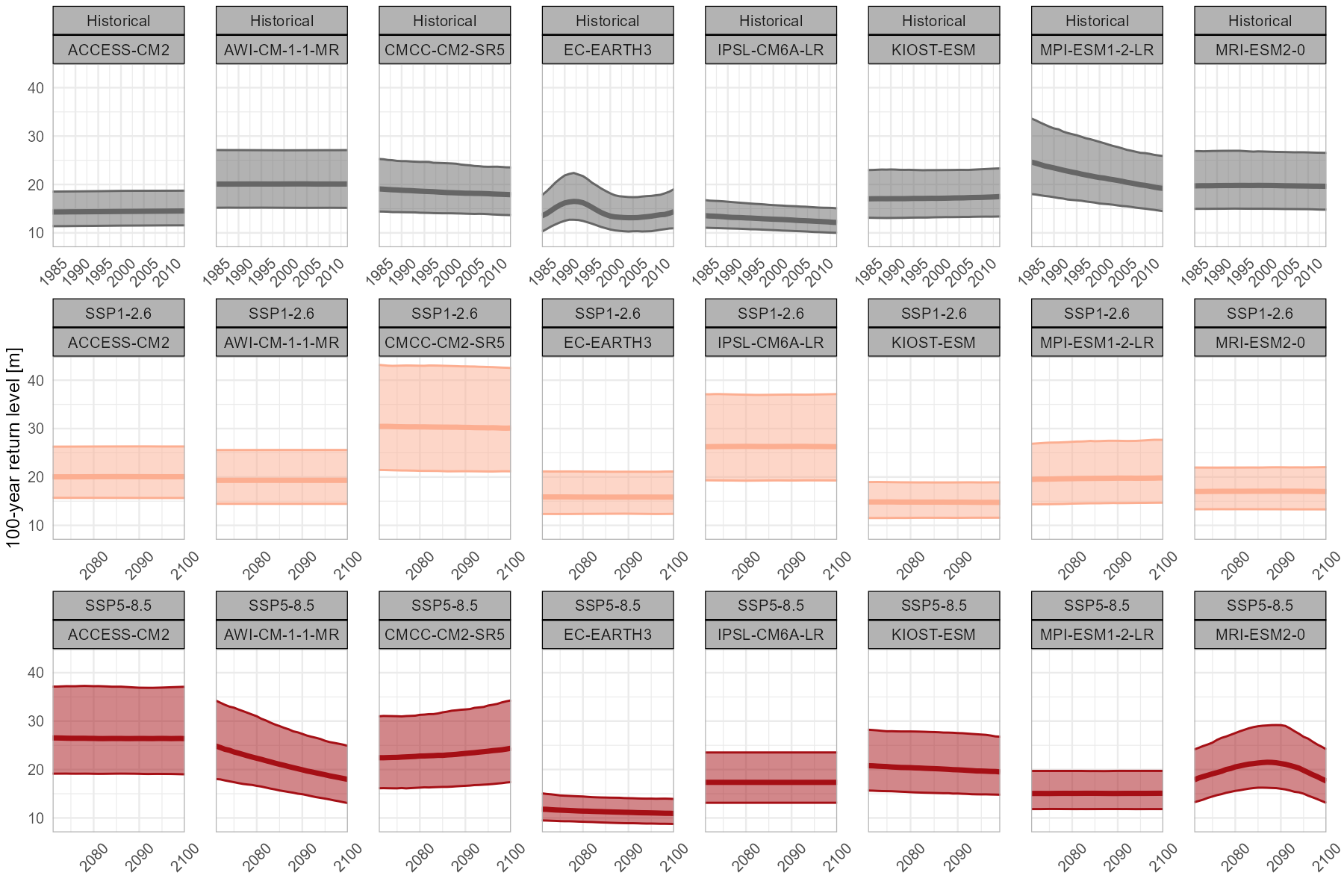}
\caption{Comparison of yearly return levels, at the West Mediterranean Representative point}
\label{fig:yearly_RL_SRP5}
\end{figure}

\begin{figure}[ht]
\includegraphics[width=\textwidth]{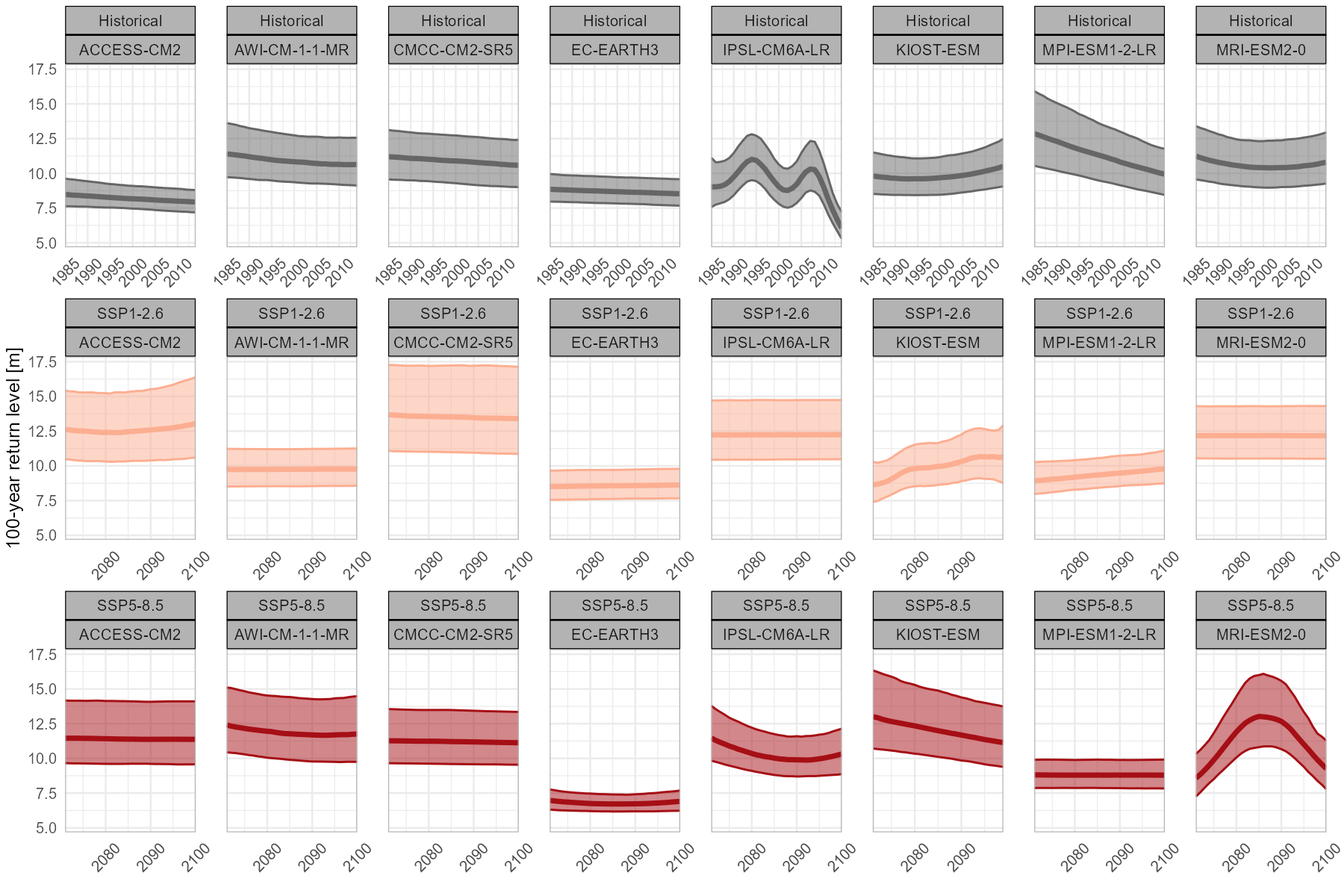}
\caption{Comparison of yearly return levels, at the East Mediterranean Representative point}
\label{fig:yearly_RL_SRP6}
\end{figure}

\FloatBarrier

\section{Lifetime design level for each SRPs}

\begin{figure}[ht]
\includegraphics[width=\textwidth]{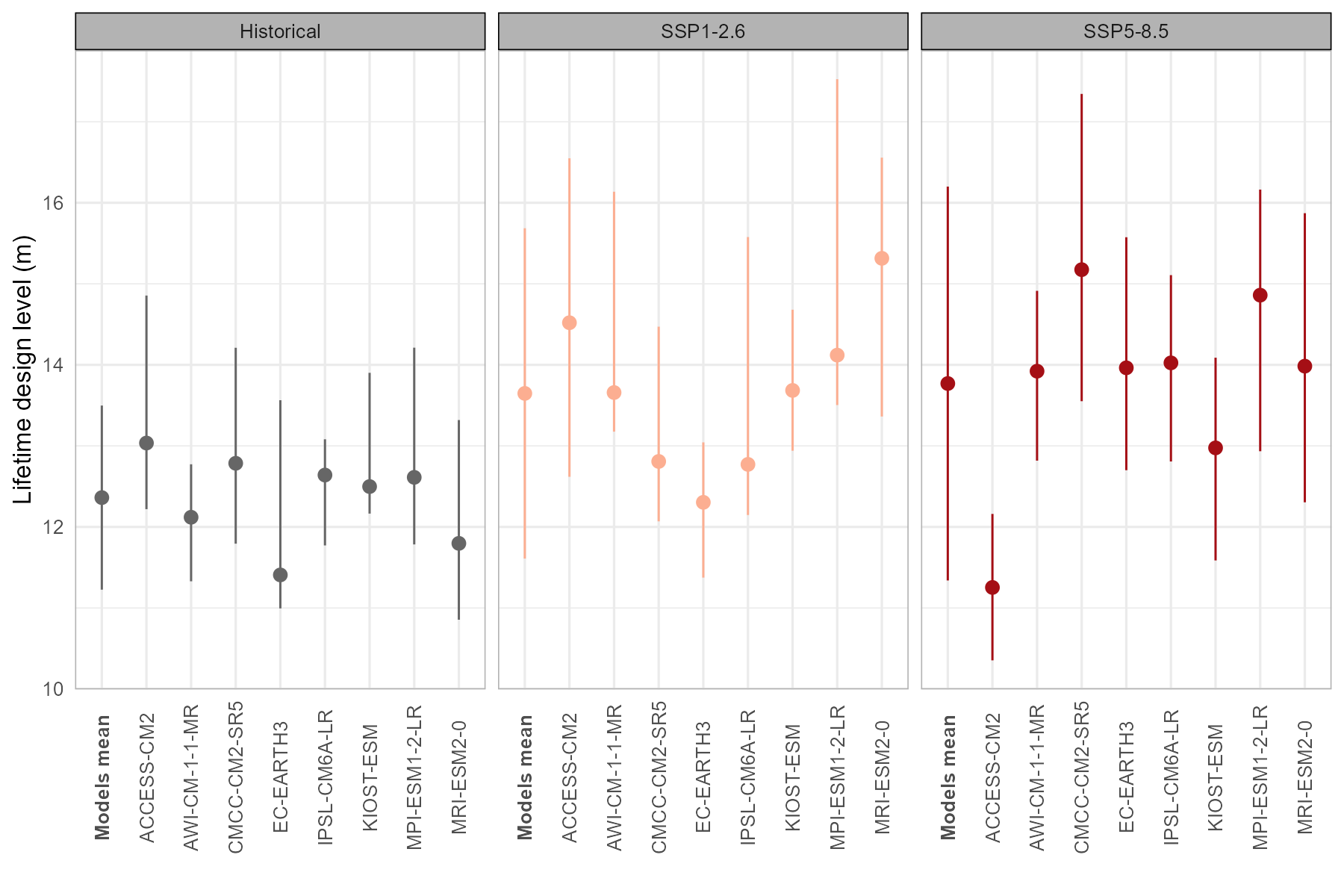}
\caption{Comparison of yearly return levels, at the North Atlantic Representative point}
\label{fig:lifetime_RL_SRP3}
\end{figure}

\begin{figure}[ht]
\includegraphics[width=\textwidth]{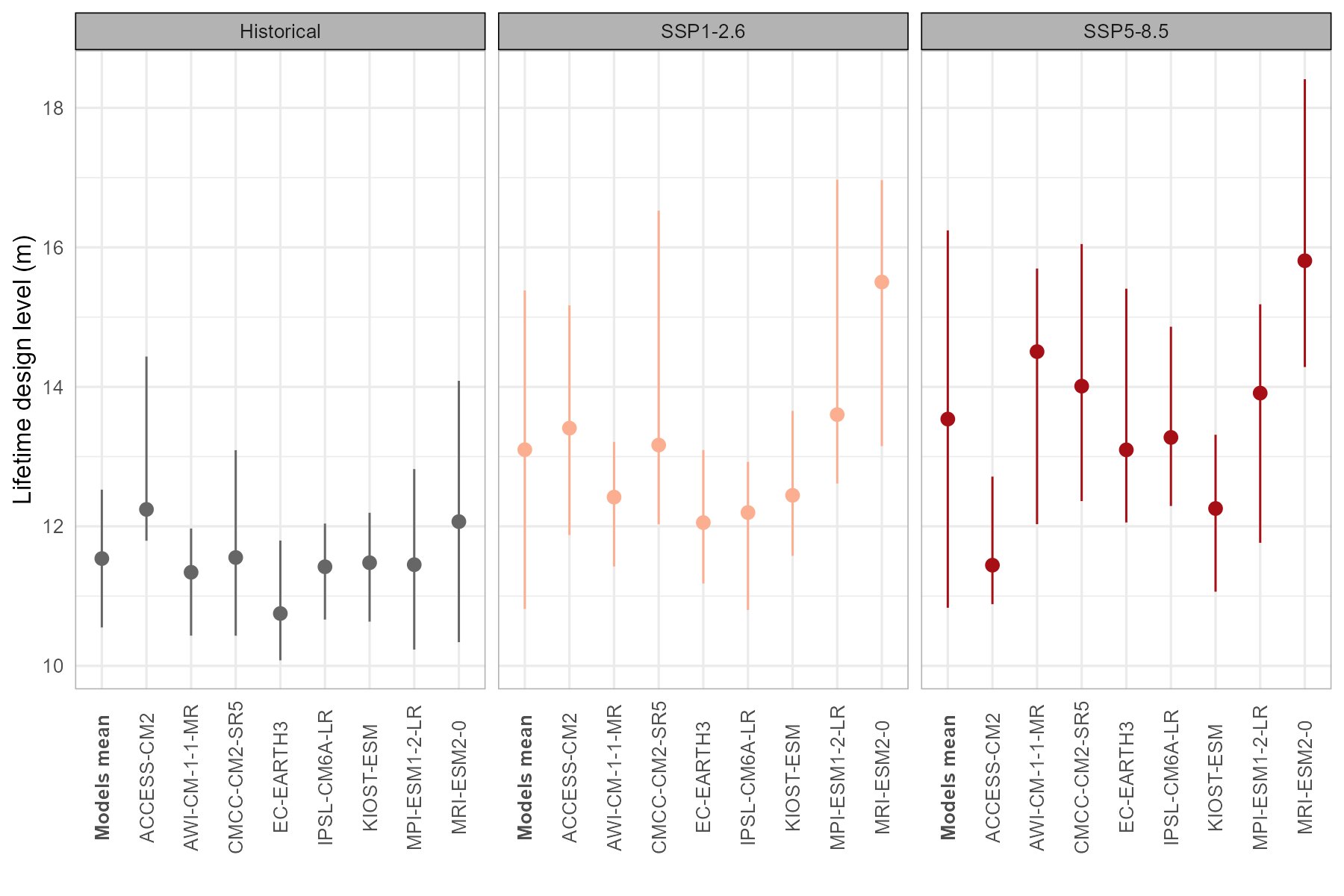}
\caption{Comparison of yearly return levels, at the South Atlantic Representative point}
\label{fig:lifetime_RL_SRP4}
\end{figure}

\begin{figure}[ht]
\includegraphics[width=\textwidth]{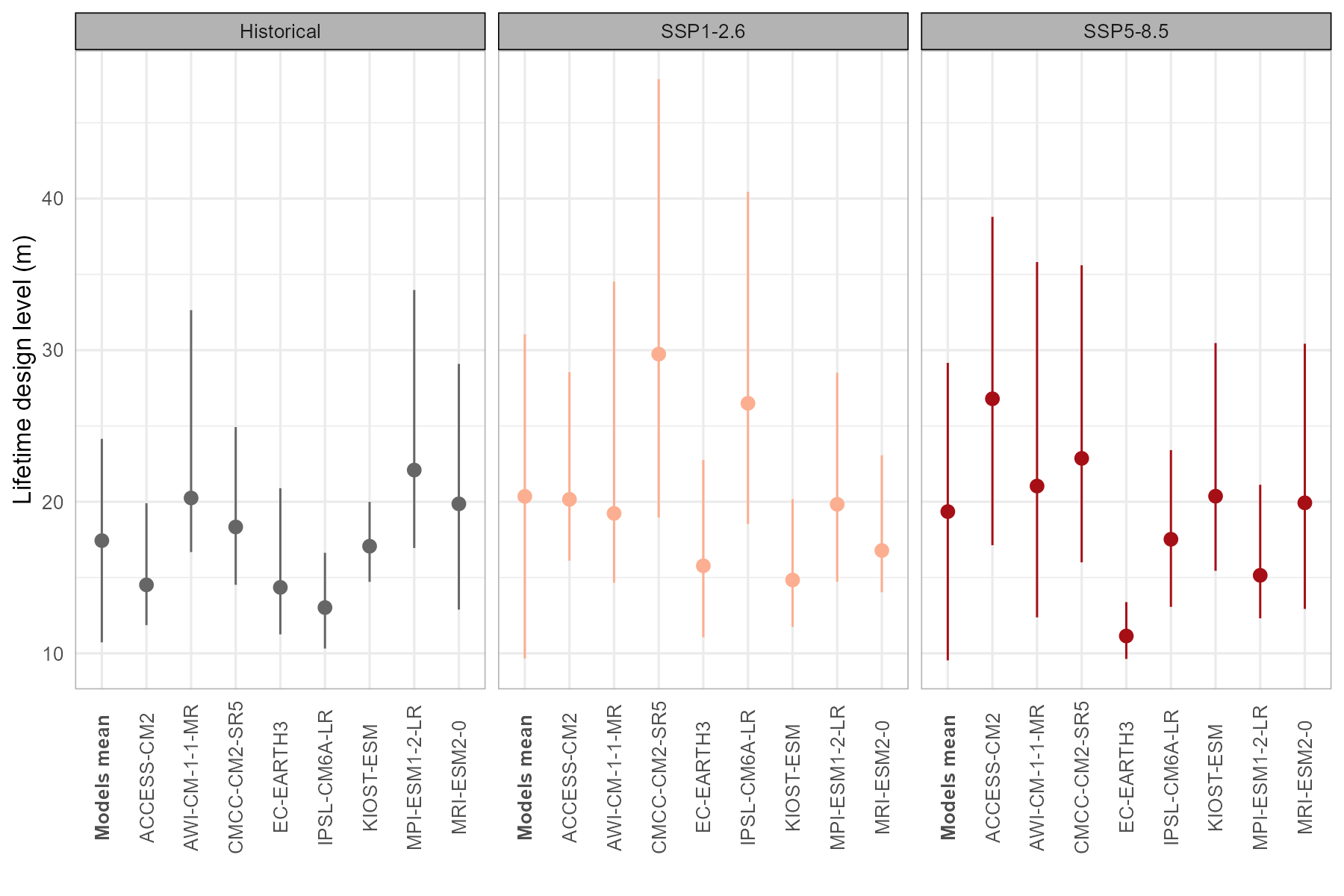}
\caption{Comparison of yearly return levels, at the West Mediterranean Representative point}
\label{fig:lifetime_RL_SRP5}
\end{figure}

\begin{figure}[ht]
\includegraphics[width=\textwidth]{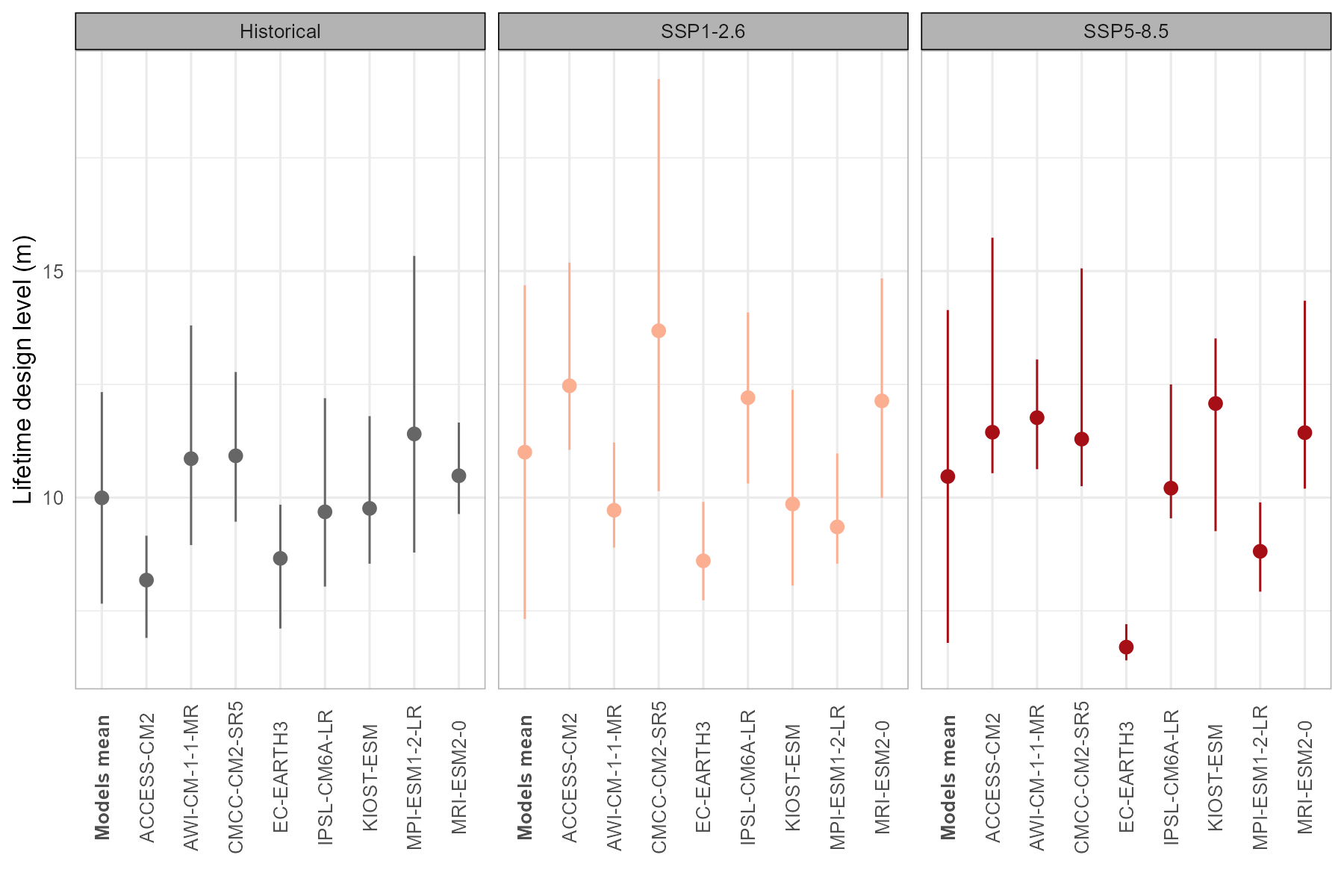}
\caption{Comparison of yearly return levels, at the East Mediterranean Representative point}
\label{fig:lifetime_RL_SRP6}
\end{figure}